\newcommand\techrep

\ifdefined\techrep
\documentclass[english]{article}
\usepackage{fullpage}
\else
\documentclass[preprint,numbers,english]{sigplanconf}
\fi
\usepackage[T1]{fontenc}
\usepackage[latin9]{inputenc}
\usepackage{color}
\usepackage{float}
\usepackage{amsthm}
\usepackage{amsmath}
\usepackage{amssymb}
\usepackage{wasysym}
\usepackage{stmaryrd}
\usepackage{graphicx}
\usepackage{subfig}

\newcommand{\isrefinedby}{\sqsubseteq}
\newcommand{\rfail}{\mathsf{fail}}
\newcommand{\impl}{\Rightarrow}
\makeatletter

\providecommand{\tabularnewline}{\\}

\theoremstyle{plain}
\newtheorem{thm}{\protect\theoremname}
  \theoremstyle{definition}
  \newtheorem{defn}{\protect\definitionname}
  \theoremstyle{plain}
  \newtheorem{lem}{\protect\lemmaname}
\newenvironment{lyxlist}[1]
{\begin{list}{}
{\settowidth{\labelwidth}{#1}
 \setlength{\leftmargin}{\labelwidth}
 \addtolength{\leftmargin}{\labelsep}
 }}
{\end{list}}
  \theoremstyle{definition}
  \newtheorem{example}{\protect\examplename}

\usepackage{hyperref}
\newcommand{\myhref}[1]{\href{#1}{\url{#1}}}

\usepackage{babel}
\providecommand{\definitionname}{Definition}
  \providecommand{\examplename}{Example}
  \providecommand{\lemmaname}{Lemma}
\providecommand{\theoremname}{Theorem}

\makeatother

\usepackage{babel}
  \providecommand{\definitionname}{Definition}
  \providecommand{\examplename}{Example}
  \providecommand{\lemmaname}{Lemma}
\providecommand{\theoremname}{Theorem}

\ifdefined\techrep
\newcommand{\authorinfo}[2]{\author{#1\\ {\normalsize #2}}}
\fi

\begin{document}

\setlength{\pdfpageheight}{\paperheight}
\setlength{\pdfpagewidth}{\paperwidth}
\ifdefined\techrep
\else
\conferenceinfo{LICS 2016}{July 5 -- 8, 2016, New York City, USA}
\copyrightyear{2016}
\copyrightdata{978-1-nnnn-nnnn-n/yy/mm}
\copyrightdoi{nnnnnnn.nnnnnnn}
\fi

\title{Towards Compositional Feedback in Non-Deterministic and
Non-Input-Receptive Systems\thanks{This work was partially supported by the Academy of Finland, the U.S. National Science Foundation (awards \#1329759 and \#1139138), and by UC Berkeley's iCyPhy Research Center (supported by IBM and United Technologies).}}

\authorinfo{Viorel Preoteasa$^{1}$ and Stavros Tripakis$^{1,2}$}{$^1$Aalto University and\ \  $^2$University of California, Berkeley}{}
\maketitle

\begin{abstract}{
Feedback is an essential composition operator in many classes 
of reactive and other systems. This paper studies feedback in 
the context of compositional theories with refinement. Such 
theories allow to reason about systems on a component-by-component 
basis, and to characterize substitutability as a refinement 
relation. Although compositional theories of feedback do exist, 
they are limited either to deterministic systems (functions)
or input-receptive systems (total relations). In this work we 
propose a compositional theory of feedback which applies to 
non-deterministic and non-input-receptive systems (e.g., partial 
relations). To achieve this, we use the semantic frameworks of 
predicate and property transformers, and relations with fail 
and unknown values. We show how to define instantaneous feedback 
for stateless systems and feedback with unit delay for stateful 
systems. Both operations preserve the refinement relation, and 
both can be applied to non-deterministic and non-input-receptive
systems.
}
\end{abstract}

\global\long\def\optionn{\mathsf{Opt}}
 \global\long\def\maximal{\mathsf{maximal}}
 \global\long\def\Fail{\mathsf{Fail}}
 \global\long\def\bool{\mathsf{bool}}

\global\long\def\Fbend{\mathsf{Fb\_end}}
 \global\long\def\Fb{\mathsf{Fb\_hide}}
 \global\long\def\Fba{\mathsf{Fb\_a}}
 \global\long\def\Fbb{\mathsf{Fb\_b}}
 \global\long\def\Fbfun{\mathsf{InstFeedbackPT}}
 \global\long\def\Fbbegin{\mathsf{Fb\_begin}}

\global\long\def\fbend{\mathsf{fb\_end}}
 \global\long\def\fb{\mathsf{fb\_hide}}
 \global\long\def\fba{\mathsf{fb\_a}}
 \global\long\def\fbb{\mathsf{fb\_b}}
 \global\long\def\fbfun{\mathsf{InstFeedback}}
 \global\long\def\fbbegin{\mathsf{fb\_begin}}

\global\long\def\ifs{\mathsf{if}}
 \global\long\def\thens{\mathsf{then}}
 \global\long\def\elses{\mathsf{else}}
 \global\long\def\addafail{\mathsf{AddFail}}
 \global\long\def\pre{\mathsf{prec}}
 \global\long\def\rel{\mathsf{rel}}
 \global\long\def\wpr{\mathsf{wp}}
 \global\long\def\nat{\mathsf{nat}}
 \global\long\def\tru{\mathsf{true}}
 \global\long\def\fals{\mathsf{false}}
 \global\long\def\divrel{\mathsf{Div}}

\global\long\def\ifstm#1#2#3{\mathsf{if}\ #1\ \mathsf{then} \ #2 \ \mathsf{else}\ #3}

\global\long\def\assume#1{[#1]}
 \global\long\def\assert#1{\{#1\}}

\global\long\def\comp{\;;\,}
 \global\long\def\feedbackmap{\mathsf{FeedbackMap}}
 \global\long\def\skipstm{\mathsf{Skip}}
 \global\long\def\feedback{\mathsf{InstFeedback}}
 \global\long\def\typeunita{\mathsf{Unit}}

\global\long\def\havocc{\mathsf{Havoc}}
 \global\long\def\fail{\mathsf{Fail}}
 \global\long\def\magic{\mathsf{Magic}}
 \global\long\def\nat{\mathsf{Nat}}
 \global\long\def\tru{\mathsf{true}}
 \global\long\def\fals{\mathsf{false}}
 \global\long\def\nxt{\mathsf{NextOld}}
 \global\long\def\itenext{\mathsf{IteNext}}

\global\long\def\L{\,\mathsf{L}\,}
 \global\long\def\nextop{\mathsf{Next}}
 \global\long\def\fusion{\mathsf{Fusion}}
 \global\long\def\always{\mathsf{Always}}
 \global\long\def\iterate{\mathsf{Iterate}}
 \global\long\def\inp{\mathsf{in}}
 \global\long\def\init{\mathsf{init}}
 \global\long\def\leads{\mathsf{\, L\,}}

\global\long\def\symfeedback{\mathsf{SymFeedback}}
 \global\long\def\trans{\mathsf{Trans}}
 \global\long\def\iteratenextomega{\mathsf{IterateNextOmega}}

\global\long\def\iterateomega{\mathsf{IterateOmega}}
 \global\long\def\skipnext{\mathsf{SkipNext}}
 \global\long\def\skiptop{\mathsf{SkipTop}}
 \global\long\def\delay{\mathsf{Delay}}
 \global\long\def\iterel{\mathsf{IterateRel}}

\global\long\def\eqtop{\mathsf{eqtop}}
 \global\long\def\idnext{\mathsf{idnext}}
 \global\long\def\adddelay{\mathsf{AddDelay}}
 \global\long\def\delayfeedback{\mathsf{DelayFeedback}}
 \global\long\def\bool{\mathsf{Bool}}
 \global\long\def\sskip{\mathsf{Skip}}
 \global\long\def\real{\mathsf{Real}}

\global\long\def\until{\;\mathsf{U}\;}
 \global\long\def\alway{\Box\,}
 \global\long\def\event{\Diamond\,}
 \global\long\def\nex{\fullmoon\,}
 \global\long\def\demonic#1#2#3{[#1\leadsto#2 \;|\; #3]}

\global\long\def\adapta{\mathsf{adapt}_{1}}
 \global\long\def\adaptb{\mathsf{adapt}_{2}}
 \global\long\def\adaptc{\mathsf{adapt}_{3}}
 \global\long\def\seqfb{\mathsf{SeqFeedback}}
 \global\long\def\monorel{\mathsf{mono}}

\section{Introduction}
\label{sec_intro}

Large and complex systems are best designed using component-based
rather than monolithic methodologies. In the component-based paradigm
a system is an assembly of interacting components. Each component
can itself be a subsystem composed from other components. Arbitrary
component hierarchies can be built in this manner, allowing to decompose
a complex system into simpler and further simpler subsystems.

Systems constantly evolve, and a key concern in a component-based
setting is \emph{substitutability}: when can we replace some component
$A$ with a new component $A'$? The issue is {\em property preservation}:
the replacement must not result in breaking essential properties of
the existing system. Moreover, we would like to avoid as much as possible
having to re-verify the new system from scratch, since verification
is an expensive process. Ideally, a simple check should suffice to
guarantee that the new system is correct, assuming that the old system
was correct.

Compositional theories with refinement provide an elegant framework
for reasoning about these problems. In such theories the following
notions are central: (a) the notion of component; (b) 
composition operators, which create new components by composing
existing ones; (c) the notion of {\em refinement}, which is a binary
relation between components. 
Typically two main theorems are provided by such a theory: 
\begin{enumerate}
\item {\em Preservation of refinement by composition}, which can be roughly
stated as: if $A'$ refines $A$ and $B'$ refines $B$, then the
composition of $A'$ and $B'$ refines the composition of $A$ and
$B$. 
\item {\em Preservation of properties by refinement}, which can be roughly
stated as: if $A'$ refines $A$ and $A$ satisfies property $\phi$,
then $A'$ also satisfies $\phi$. 
\end{enumerate}
To see how the above two theorems can be used, consider again the
scenario above where $A$ is replaced by $A'$. We can avoid re-verifying
the new system by trying instead to show that $A'$ refines $A$.
If this is true, then by preservation of refinement by composition
it follows that the new system (with $A'$) refines the old system
(with $A$). (Here we also implicitly use that refinement is reflexive,
which is typically true.) Moreover, by preservation of properties
by refinement it follows that if property $\phi$ holds in the old
system then it will also hold in the new system. In this way, we can
avoid checking $\phi$ directly on the new system, which can be expensive
when the system is large. Instead, we must show that $A'$ refines
$A$, which is an easier verification problem, as it deals with only
two components.

In this paper we look at components as blocks with inputs and outputs.
In such a setting, there are three basic composition operators (see
Figure~\ref{fig_compos}): 
\begin{figure}
\centering{}\includegraphics{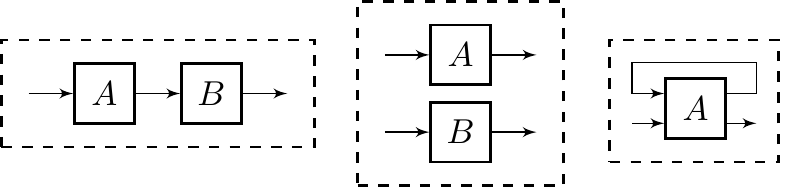}\protect\protect\caption{Composition operators: serial composition (left); parallel composition
(middle); feedback composition (right).\label{fig_compos}}
\end{figure}
\begin{itemize}
\item {\em Serial composition}, where the output of one component is
connected to the input of another component. 
\item {\em Parallel composition}, where two components are ``stacked
on top of each other'' without any connections, but encapsulated
into a single component. 
\item {\em Feedback composition}, where the output of a component is
connected into one of its inputs. 
\end{itemize}
\mbox{}\\[-3.7ex]
Semantically, we want our components to be able to model systems which
are both {\em non-deterministic} and {\em non-input-receptive}.
Non-determinism means that the system may produce more than one output
for a given input. Non-determinism is well established as a powerful
modeling mechanism which allows abstracting from unnecessary details
among other things. Non-input-receptiveness means that the system
may declare some input values to be {\em illegal} (always, or at
certain times). For example, a component computing the square root
of a real number may declare that negative inputs are illegal. Note
that this is different from stating that if the input is negative
then the output can be arbitrary. The latter corresponds logically
to an implication: $x\ge0\impl y=\sqrt{x}$ (here $x$ denotes the input
and $y$ the output); whereas the former corresponds to a conjunction:
$x\ge0\land y=\sqrt{x}$. Although less common in the domains of formal
methods and verification, non-input-receptive systems are common in
programming languages and type theory, where they provide the basis
for type checking. Type checking is a key feature of many languages
and can be seen as a ``lightweight'' verification mechanism as the
programmer does not have to provide a formal specification of her
program (the program only needs to type check). In our setting, non-input-receptiveness
is essential in order to be able to model {\em compatibility} of
components during composition. For an extensive discussion of this
point, see~\cite{AlfaroGamesOpenSystems04,TripakisTOPLAS2011}.

In this setting, parallel composition is relatively straightforward
to define and can be logically seen as taking the conjunction of the
specifications of the two components. Serial composition is a bit
trickier and logically involves $\forall$-$\exists$ quantification,
which essentially states that any output of the upstream component
is a legal input for the downstream component (i.e., there exists
for it an output of the downstream component). This {\em game theoretic}
semantics is necessary when the components are both non-deterministic
and non-input-receptive, while it collapses to the standard semantics
(composition of functions or relations, or logical conjunction) when
components are either deterministic or input-receptive~\cite{AlfaroGamesOpenSystems04,AlfaroHenzingerFSE01,TripakisTOPLAS2011}.
Both parallel and serial composition as defined above are compositional
in the sense that they preserve refinement.

Compositional theories of feedback do exist, but they are limited to
either deterministic systems, or to input-receptive systems. To our
knowledge, compositional feedback for both non-deterministic and non-input-receptive
systems has so far remained elusive. The main difficulty has been
to come up with a definition of feedback which preserves refinement.
That is, if $A'$ refines $A$, then $\kappa(A')$ refines $\kappa(A)$,
where $\kappa(\cdot)$ is the feedback composition operator. An extensive
account of some of the difficulties and failed attempts is given in~\cite{tripakisShaverFPS2014},
for the framework of \emph{relational interfaces}~\cite{TripakisTOPLAS2011}.

To illustrate some
of the earlier problems and how they are handled in the new framework,
consider the following example, first given in~\cite{DoyenHJP08}.
Let $A$ be a component with two input variables $u,x$, one output
variable $v$, and input-output relation $\tru$, which means that
$A$ accepts any input value, and given an input value, it can return
any output (because $\tru$ is satisfied by any $(u,x,v)$ triple).
Now let $A'$ be another component with inputs $u,x$, output $v$,
and relation $u\neq v$. This means that $A'$ accepts any input $(u,x)$,
and given such $u$, returns some $v$ different from $u$. In the
relational interface framework (as well as in our framework), $A'$
refines $A$. This is because $A'$ has the same legal inputs as $A$
(in fact, every input is legal for both) and $A'$ is ``more deterministic''
than $A$.

Now, suppose we apply feedback to both $A$ and $A'$, in particular,
we connect output $v$ to input $u$. Denote the resulting components
by $B$ and $B'$, respectively. Both $B$ and $B'$ have a single
input, $x$ ($u$ is no longer a free input, since it has been connected
to $v$). What should the relations of $B$ and $B'$ be? Note that
we want $B'$ to refine $B$, otherwise refinement is not preserved
by feedback. The straightforward way to define feedback is to add
the constraint $u=v$, since $v$ is connected to $u$, essentially
becoming the same ``wire''. Doing so, we obtain $\tru\land u=v$,
i.e., $u=v$, as the relation of $B$, and $u\neq v\land u=v$, i.e.,
$\fals$, as the relation of $B'$. This is problematic, since $\fals$
does not refine $u=v$. Indeed, according to the game-theoretic semantics,
refinement is \emph{not} implication, because inputs and outputs must
be treated differently. In our example, any input $x$ is legal for
$B$, whereas no input $x$ is legal for $B'$, which means that $B'$
cannot replace $B$, i.e., does not refine it.

In this paper, we attack the problem of feedback afresh. We adopt
the frameworks of {\em refinement calculus} (RC)~\cite{back-wright-98}
and {\em refinement calculus of reactive systems} (RCRS)~\cite{preoteasa:tripakis:2014}.
RC is a classic compositional framework for reasoning about sequential
programs, while RCRS is an extension of RC to reason about reactive
systems. In RC, programs are modeled as (monotonic) \emph{predicate
transformers}, i.e., functions from sets of post-states to sets of
pre-states. In RCRS, systems are modeled as (monotonic) \emph{property
transformers}, which are functions that transform sets of output infinite
sequences into sets of input infinite sequences. Both predicate and
property transformers are powerful mathematical tools, able to capture
both non-deterministic and non-input-receptive systems. In fact, they
are more powerful than partial relations, which is the semantic basis
of relational interfaces~\cite{TripakisTOPLAS2011}. Therefore one can hope
that some of the difficulties in obtaining a compositional definition
of feedback in the relational framework can be overcome in the refinement
calculi frameworks.

We first consider {\em stateless} systems and {\em instantaneous feedback},
i.e., feedback in a single step (Section~\ref{sec:inst-feedback}).
We model stateless systems using predicate transformers, where instead
of a post/pre-state interpretation, we use an output/input interpretation.
We often employ {\emph{strictly conjunctive}} predicate transformers,
which correspond to input-output relations over a domain extended
with a special value {\em fail} (denoted $\bullet$), used to model
illegal inputs (those return $\bullet$ as output). We further extend
these relations with an additional {\em unknown} value (denoted
$\bot$). The unknown value is used to define feedback by means of
a fix-point iteration, starting with unknown and potentially converging
to known values. The idea to use unknown values and fix-points comes
from methods for dealing with instantaneous feedback in deterministic
and input-receptive systems (i.e., total functions)~\cite{Malik94,Berry99,EdwardsLee03}.
In our case systems are non-deterministic and also may reject some
inputs. We therefore need to generalize the fix-point method to relations
with fail and unknown. 
The precise definition is involved (Def.~\ref{def:inst-feedback}),
but roughly speaking, for a given input we explore all possible computation
paths, including those in which the feedback input stabilizes to a
known or unknown value, as well as those in which either the relation
returns $\bullet$ or the path does not stabilize.
We can then consider an input to the post-feedback system as legal if
no execution path fails, and those which stabilize reach a known value.

This instantaneous feedback operator preserves refinement.
In the example discussed above, the new definition results
in both $B$ and $B'$ being the component $\fail$ which rejects
all inputs, since the above procedure does not stabilize in neither
the case of $A$ nor $A'$. This indicates that the two feedback operations
are illegal, which is similar to discovering an incompatibility during,
say, serial composition.

Relations with fail and unknown have the same modeling power as strictly
conjunctive predicate transformers over input/output values with unknowns.
We use the two models alternatively, as each is best suited to obtain
certain results. In particular, we define the instantaneus feedback
operator using relations with fail and unknowns, but we prove that
the operator preserves refinement using their predicate transformer
representation.

We next turn to the case of {\em stateful} systems.
Syntactically, we model these as {\em symbolic transition systems},
i.e., relations on input, output, current state, and next state variables.
For instance, a simple counter can be modeled by the relation $u'=u+1$, 
where $u$ denotes the current state and $u'$ the next state.
Semantically, we use predicate transformers on the same variables~\cite{preoteasa:tripakis:2014}.
We then define {\em feedback with unit-delay} for such predicate transformers,
which corresponds to connecting $u'$ to $u$ via a {\em unit-delay} component.
A unit-delay outputs at step $k$ its input at the previous step $k-1$.
Thus, feedback with unit-delay ensures that the next state $u'$ becomes 
the current state in the next step.
The feedback with unit-delay operator is defined
in Section~\ref{sec:feedback-unit-delay}. That definition is aided by iteration
operators over property transformers defined in Section~\ref{sec:iteration}.

Combining the two feedback operators generalizes block-diagram languages like
Simulink, which are not able to handle non-deterministic and non-input-receptive
components. 

Our treatment in this paper is mostly semantic.
However, following~\cite{preoteasa:tripakis:2014},
we can use syntactic objects such as symbolic transition systems
to represent predicate and property transformers. 
For example, the formula $u'=u+1$ represents a counter, as mentioned above;
the formula $x\ge0\land y=\sqrt{x}$, where $x$ is input and $y$ is output,
represents a relation with fail and also a predicate transformer.
We can also use {\em temporal logic} (e.g., LTL~\cite{pnueli:1977})
to express properties over infinite sequences of input-output values,
including liveness properties.
For example, a system
specified by the LTL formula $\alway(x\impl\event y)$
ensures that if the input $x$ is infinitely often true, so will be
the output $y$. 
Using these representations, most of
the constructs presented in this paper can actually be computed as
first-order formulas, or as quantified LTL formulas.

All the results presented in this paper have been formalized and proved
in the Isabelle theorem prover \cite{nipkow-paulson-wenzel-02}. The
Isabelle code is available from
\url{http://rcrs.cs.aalto.fi}.
\ifdefined\techrep
\else
An extended version of this paper is also available~\cite{preoteasa:tripakis:2015}.
\fi

\section{Related Work}

Our idea to use relations with an unknown value is inspired from fix-point based
methods such as \emph{constructive semantics}, which are used
to reason about instantaneous feedback in cyclic combinational circuits
and synchronous languages~\cite{Malik94,Berry99,EdwardsLee03}.
These methods deal with monotonic functions over the flat CPO with
unknown as the least element. Because they assume total functions,
they are limited to deterministic and input-receptive systems.
Refinement is also absent from these frameworks.

Fix-point based methods have also been used to reason about non-instantaneous
feedback, in systems such as Kahn Process Networks~\cite{Kahn74},
where processes are modeled as continuous functions on streams. Again,
these are total functions, and therefore deterministic and input-receptive.
Non-deterministic extensions of process networks 
exist~\cite{Jonsson94fully} but they do not include a notion of refinement.

Frameworks which allow non-determinism and also include refinement
are I/O automata~\cite{LynchTuttle89}, reactive modules~\cite{AlurHenzingerFMSD99},
and Focus~\cite{BroyStolen01}. 
All three frameworks assume that components are input-receptive. Note
that in the example given in the introduction, although components
$A$ and $A'$ are input-receptive, and so is $B$, $B'$ is not.
This shows that the straightforward definition of feedback does not
preserve input-receptiveness. It is unclear how this example could
be handled in the above frameworks.

Interface automata \cite{AlfaroHenzingerFSE01} allow non-deterministic
as well as non-input-receptive specifications. Composition in this
framework is asynchronous parallel composition with input-output label
synchronization, and therefore difficult to directly compare with
feedback composition in our framework. Also, interface automata are
limited to safety whereas our framework can also express liveness
properties.

Checking compatibility within the framework of functional
programing languages, can be done using type-checking as in the Standard
ML language \cite{Milner:1990}, or using more advanced techniques
as Refinement Types for ML \cite{Freeman:Pfenning:1991}, Dependent
Types \cite{Xi:Pfenning:1999}, and Liquid Types \cite{Rondon:Kawaguci:Ranjit:2008}.
These are techniques based on subtypes and dependent-types that allow
checking at compile time of invariants about the computed values of
the program. These techniques are in general automated, therefore
they are applicable only to certain classes of decidable problems.
Our compatibility check for system compositions is more general, and
not necessarily decidable. In our case checking compatibility
of system compositions is reduced to checking satisfiability of formulas.
If the underlying logic of these formulas is decidable,
then we can have an automatic compatibility test.

In \cite{millen} a model for synchronous nondeterministic systems is used
for studying preservation of security properties by composition of systems,
including delayed feedback. However this model does not treat non-input
receptive systems.

\section{Background}

We use higher order logic as implemented by the Isabelle
theorem prover. Capital letters $X,Y,\ldots$ denote types and
small letters denote elements of these types $x\in X$. 
$\bool$ denotes the type of Boolean values $\tru$ and $\fals$, $\nat$
the natural numbers, and $\real$ the reals.
We use $\land$, $\lor$, $\impl$, and $\neg$ for the Boolean
operations.

If $X$ and $Y$ are types, then $X\to Y$ denotes the type of function
from $X$ to $Y$. For function application we use 
\ifdefined\techrep
a \emph{dot notation}.
If $f:X\to Y$ is a function, and $x\in X$, then $f.x$ is the function
$f$ applied to $x$.
\else
$f.x$ instead of $f(x)$.
\fi
The function type constructor associates to
right, and correspondingly the function application associates to
left. That is $X\to Y\to Z$ is the same as $X\to(Y\to Z)$ and $f.x.y$
is the same as $(f.x).y$. We also use lambda abstraction to construct
functions. For example if $x+2\cdot y\in\nat$ is natural expression,
then $(\lambda x,y:x+2\cdot y):\nat\to\nat\to\nat$ is the function
that maps $x$ and $y$ into $x+2\cdot y$. 
\ifdefined\techrep
We use the notation $X\times Y$
for the Cartesian product of $X$ and $Y$ and if $x\in X$ and $y\in Y$,
then $(x,y)$ denotes the pair of $x$ and $y$ from $X\times Y$.
\else
$X\times Y$ denotes the Cartesian product of $X$ and $Y$.
\fi

Predicates are functions with one or more arguments returning Booleans
(e.g., a predicate with two arguments has signature $X\to Y\to\bool$),
and relations are predicates with at least two arguments. For a relation
$r:X\to Y\to\bool$ we denote by $\inp.r:X\to\bool$ the predicate
given by $\inp.r.x=(\exists y:r.x.y)$. If $r$ is a relation
with more than two arguments, then $\inp.r$ is defined similarly
by quantifying over the last argument: $\inp.r.x.y.z=(\exists u:r.x.y.z.u)$.

We extend point-wise all operations on Booleans to operations on predicates
and relations. For example if $r,r':X\to Y\to\bool$ are two relations
then $r\land r'$ and $r\lor r'$ are the relations given by $(r\land r').x.y=r.x.y\land r'.x.y$
and $(r\lor r').x.y=r.x.y\lor r.x.y$, and $(r\le r')=(\forall x,y:r.x.y\impl r'.x.y)$.
We use $\bot$ and $\top$ as the smallest and greatest predicates
(relations): $\bot.x=\fals$ and $\top.x=\tru$.

For relations $r:X\to Y\to\bool$ and $r':Y\to Z\to\bool$, we denote
by $r\circ r'$ the \emph{relational composition} given by $(r\circ r').x.z=(\exists y:r.x.y\land r'.y.z)$.
For a relation $r:X\to Y\to\bool$ we denote by $r^{-1}:Y\to X\to\bool$
the \emph{inverse} of $r$ ($r^{-1}.y.x\Leftrightarrow r.x.y$). A
relation $r$ is \emph{total} (sometimes called \emph{left-total})
if for all $x$ there is $y$ such that $r.x.y$.

For reactive systems we need \emph{infinite sequences} (or \emph{traces})
of values. Formally, if $X$ is a type of values, then a trace from
$X$ is an element $\mathbf{x}\in X^{\omega}=(\nat\to X)$. To make
the reading easier, we denote infinite sequences by bold face letters
$\mathbf{x},\mathbf{y},\ldots$ and values by normal italic small
letters $x,y,\ldots$. When, for example, both $x$ and $\mathbf{x}$
occur in some formula, we use the convention that the elements of
$\mathbf{x}$ have the same type as $x$. For $\mathbf{x}\in X^\omega$,
$\mathbf{x}_{i}=\mathbf{x}.i$, and $\mathbf{x}^{i}\in X^{\omega}$
is given by $\mathbf{x}_{j}^{i}=\mathbf{x}^{i}.j=\mathbf{x}_{i+j}$.
We consider a pair of traces $(\mathbf{x},\mathbf{y})$ as being the
same as the trace of pairs $(\lambda i:(\mathbf{x}_{i},\mathbf{y}{}_{i}))$.

\subsection{Linear temporal logic}

Linear temporal logic (LTL) \cite{pnueli:1977} is a logic used for
specifying properties of reactive systems. In this paper we use a
semantic (algebraic) version of LTL as in \cite{preoteasa:tripakis:2014}.
This logic is formalized in Isabelle, and the details of this formalization
are available in \cite{preoteasa:2014}. An LTL formula is a predicate
on traces and the temporal operators are functions mapping predicates
to predicates. We call predicates over traces (i.e., sets of traces)
\emph{properties}.

If $p,\; q\in X^{\omega}\to\bool$ are properties, then \emph{always}
$p$, \emph{next} $p$, and $p$ \emph{until} $q$ are also properties
and they are denoted by $\Box\, p$, $\nex p$, and $p\until q$ respectively.
The property $\Box\, p$ is true in $\mathbf{x}$ if $p$ is true
at all time points in $\mathbf{x}$, $\nex p$ is true in $\mathbf{x}$
if $p$ is true at the next time point in $\mathbf{x}$, and $p\until q$
is true in $\mathbf{x}$ if there is some time in $\mathbf{x}$ when
$q$ is true, and until then $p$ is true. Formally we have: 
\[
\begin{array}{l}
(\Box\, p).\mathbf{x}=(\forall n:p.\mathbf{x}^{n}),\ \ (\nex p).\mathbf{x}=p.\mathbf{x}^{1}\\
(p\until q).\mathbf{x}=(\exists n:(\forall i<n:p.\mathbf{x}^{i})\land q.\mathbf{x}^{n})
\end{array}
\]
In addition we define the operator $p\leads q=\neg(p\until\neg q)$.
Intuitively, $p\leads q$ holds if, whenever $p$ holds continuously
up to step $n-1$, then $q$ must hold at step $n$.

If $r$ is a relation on traces we define the temporal operators similarly.
For example $(\alway r).\mathbf{x}.\mathbf{y}.\mathbf{z}=(\forall n:r.\mathbf{x}^{n}.\mathbf{y}^{n}.\mathbf{z}^{n})$.
If $r$ is a relation on values, then we extend it to traces by $r.\mathbf{x}.\mathbf{y}.\mathbf{z}=r.\mathbf{x}_{0}.\mathbf{y}_{0}.\mathbf{z}_{0}$.

\subsection{\label{sub:refinement}Refinement calculus of reactive systems}

In refinement calculus programs are modeled as \emph{monotonic predicate
transformers}~\cite{back-wright-98}.
A program $S$ with input type $X$ and output type
$Y$ is modeled as a function from $(Y\to\bool)\to(X\to\bool)$ with
a weakest precondition interpretation. If $q:Y\to\bool$ is a post-condition
(set of final states), then $S.q$ is the set of all initial states
from which the program always terminates and it terminates in a state
from $q$. We model reactive systems as \emph{monotonic property transformers},
also with a weakest precondition interpretation~\cite{preoteasa:tripakis:2014}. A reactive system
$S$ with input type $X$ and output type $Y$ is formally modeled
as a monotonic function from $(Y^{\omega}\to\bool)\to(X^{\omega}\to\bool)$.
If $q:Y^{\omega}\to\bool$ is a post-property (a set of output traces),
then $S.q$ is the set of input traces for which the execution of
$S$ does not fail and it produces an output trace in $q$. 
The remaining of this section applies equally to both monotonic predicate
transformers and monotonic property transformers, and we will call
them simply monotonic transformers.

The point-wise extension of the Boolean operations to predicates,
and then to monotonic transformers gives us a \emph{complete lattice}
with the operations $\sqsubseteq$, $\sqcap,\;\sqcup,\ \bot,\;\top$
(observe that $\sqcap$ and $\sqcup$ preserve monotonicity). All
these lattice operations are also meaningful as operations on programs
and reactive systems. The order of this lattice ($S\sqsubseteq T\Leftrightarrow(\forall q:S.q\le T.q)$)
is the \emph{refinement relation}. If $S\sqsubseteq T$, then we say
that $T$ \emph{refines} $S$ (or $S$ is \emph{refined} by $T$).
If $T$ refines $S$ then $T$ can replace $S$ in any context.

The operations $\sqcap$ and $\sqcup$ model \emph{demonic} and \emph{angelic}
\emph{non-determinism} or \emph{choice}. The interpretation of $\sqcap$
is that the system $S\sqcap T$ is correct (i.e., satisfies its specification)
if both $S$ and $T$ are correct. On the other hand $S\sqcup T$
is correct if one of $S$ and $T$ are correct. Thus, $\sqcap$ and
$\sqcup$ represent uncontrollable and controllable non-determinism,
respectively.

We denote the bottom and the top of the lattice of monotonic transformers
by $\fail$ and $\magic$ respectively. $\fail$ is refined by any
monotonic transformer and $\magic$ refines every monotonic transformer.
$\fail$ does not guarantee any property. For any $q$, $\fail.q=\bot$,
i.e., there is no input (sequence) for which $\fail$ will produce
an output (sequence) from $q$. On the other hand $\magic$ can establish
any property $q$ ($\magic.q=\top$). The problem with $\magic$ is
that it cannot be implemented.

\emph{Serial composition} of two systems modeled as transformers $S$
and $T$ is simply the functional composition of $S$ and $T$ ($S\circ T$).
For monotonic transformers we denote this composition by $S\comp T$
($(S\comp T).q=S.(T.q)$). To be able to compose $S$ and $T$, the
type of the output of $S$ must be the same as the type of the input
of $T$.

The system $\sskip$ defined by $(\forall q:\sskip.q=q)$ is the neutral
element for serial composition: 
\[
\sskip\comp S=S\comp\sskip=S.
\]

We now define two special types of monotonic transformers which will
be used to form more general property transformers by composition.
For predicate $p:X\to\bool$ and relation $r:X\to Y\to\bool$
we define the \emph{assert transformer} $\assert p:(X\to\bool)\to(X\to\bool)$,
and the \emph{demonic update transformer} $\assume r:(Y\to\bool)\to(X\to\bool)$
as follows: 
\[
\assert p.q = (p\land q),\ \ \ \ \ \assume r.q.x=(\forall y:r.x.y\impl q.y)
\]
The assert system $\assert p$ when executed from a sequence $x$,
behaves as skip when $p.x$ is true, and it fails otherwise. The demonic
update statement $\assume r$ establishes a post condition $q$ when
starting in $x$ if all $y$ with $r.x.y$ are in $q$. If $r=\bot$,
then $[r]=[\bot]=\magic$.

If $R$ is an expression in $x$ and $y$, then $\demonic xyR=\assume{\lambda x,y:R}$.
For example if $R$ is $z=x+y$, then $\demonic{x,\; y}z{z=x+y}=\assume{\lambda(x,\; y),\; z:z=x+y}$
is the system which produces in the output $z$ the value $x+y$ where
$x$ and $y$ are the inputs. If $e$ is an expression in $x$, then
$[x\leadsto e]=\assume{x\leadsto y\;|\; y=e}$, where $y$ is a new
variable different of $x$ and which does not occur free in $e$.
For assert statements we introduce similar notation. If $P$ is an
expression in $x$ then $\assert{x\;|\; P}=\assert{\lambda x:P}$.
If $P$ is $x\le y$, then $\assert{x,\; y\;|\; x\le y}=\assert{\lambda(x,y):x\le y}$.
The variables $x$ and $y$ are bounded in $\demonic xyR$ and $\assert{x\;|\; P}$.

The \emph{havoc monotonic transformer} is the most non-deterministic
statement $\havocc=[x\leadsto y\;|\;\tru]$ and it assigns some arbitrary
value to the output $y$.

In RCRS we can model (global) specifications of reactive systems, e.g.,
using LTL as a specification language, as well as concrete implementations,
e.g., using symbolic transition systems (relations on input, output,
state, and next-state variables)~\cite{preoteasa:tripakis:2014}.
A global specification may define what the output of a system is when
the input is an entire sequence $\mathbf{x}$.  
A global specification may also impose conditions on the entire input
sequence $\mathbf{x}$, including liveness conditions.
For example, the formula $\alway\event x$ asserts that the input
\emph{must} be true infinitely often. An implementation describes
how the system works step-by-step. For instance, an implementation
may start with input $\mathbf{x}_{0}$ and some initial internal state
$\mathbf{u}_{0}$, compute the output $\mathbf{y}_{0}$ and a new
internal state $\mathbf{u}_{1}$, and then use $\mathbf{x}_{1}$ and
$\mathbf{u}_{1}$ to compute $\mathbf{y_{1}}$ and $\mathbf{u}_{2}$,
and so on.

\begin{example}[System modeling with property and predicate transformers]
We can model a global reactive system $\mathit{Sum}$ that for
every input sequence $\mathbf{x}\in\nat^{\omega}$ computes the sequence
$sum.\mathbf{x}\in\nat^{\omega}$, where $sum.\mathbf{x}.n=\mathbf{x}_{0}+\mathbf{x}_{1}+\ldots+\mathbf{x}_{n-1}$.
We model $\mathit{Sum}$ as a monotonic property transformer: 
\[
\mathit{Sum}=[\mathbf{x}\leadsto sum.\mathbf{x}]
\]
We can also model a step of an implementation of $\mathit{Sum}$ as a predicate
transformer:
\begin{eqnarray}
\label{eqstepsum}
\mathit{StepSum}=[u,x\leadsto u',y\;|\; u'=u+x\land y=u]
\end{eqnarray}
The relation $u'=u+x\land y=u$ describes a symbolic transition system:
$u$ is the current state, $u'$ the next state, $x$ is the
input, and $y$ the output. $\mathit{StepSum}$ works
iteratively starting at initial state $\mathbf{u}_{0}$, and the first
element of the input $\mathbf{x}$, and assuming that $\mathbf{u}_{0}=0$,
it computes $\mathbf{u}_{1}=\mathbf{u}_{0}+\mathbf{x}_{0}=\mathbf{x}_{0}$
and $\mathbf{y}_{0}=\mathbf{u}_{0}=0=sum.\mathbf{x}.0$, then using
$\mathbf{u}_{1}$ and second element $\mathbf{x}_{1}$ of $\mathbf{x}$,
it computes $\mathbf{u}_{2}=\mathbf{u}_{1}+\mathbf{x}_{1}=\mathbf{x}_{0}+\mathbf{x}_{1}$
and $\mathbf{y}_{1}=\mathbf{u}_{1}=\mathbf{x}_{0}=sum.\mathbf{x}.1$,
and so on. We will show later in Example \ref{ex:sum:delay} how to obtain the $\mathit{Sum}$
property transformer by connecting in $\mathit{StepSum}$ the output $u'$
to the input $u$ in feedback with a unit delay.
\end{example}

The assert and demonic choice predicate transformers satisfy the following
properties \cite{back-wright-98}: 
\begin{thm}
\label{thm:spec-comp-refin}For $p,p'$ and $r,r'$ of appropriate
types we have\end{thm}
\begin{enumerate}
\item $\assert p\comp\assume r\comp\assert{p'}\comp\assume{r'}=\{x\;|\; p.x\land\forall y:r.x.y\impl p'.y\}\comp\assume{r\circ r'}$ 
\item $\assert p\comp\assume r\sqsubseteq\assert{p'}\comp\assume{r'}\Leftrightarrow(\forall x:(p.x\impl p'.x)\land(\forall y:p.x\land r'.x.y\impl r.x.y))$ 
\end{enumerate}
\ifdefined\techrep
For a monotonic predicate transformer $S$, the \emph{strong} and\emph{
weak iteration} of $S$, denoted $S^{\omega}$ and $S^{*}$, are the
least and greatest fix-points of $(\lambda X:\sskip\sqcap S\comp X)$,
respectively. These fix-points exist by the Knaster-Tarski
fix-point theorem for monotonic functions on complete lattices~\cite{tarski-55}.
When $S^{*}$
is executed, $S$ is repeated a demonically chosen finite
number of times. $S^{\omega}$ is similar, but if $S$ can be executed
indefinitely then $S^{\omega}$ fails. $S^{\omega}$ can be used to
give semantics to the while statement suitable for total correctness
reasoning, while $S^{*}$ can be used to give semantics to the while
statement suitable for partial correctness reasoning. In this paper
we will use $S^{*}$ to give an alternative definition for the instantaneous
feedback operator.
\fi

\subsection{Product and fusion of monotonic transformers}
For two predicates $p$ and $q$ their \emph{product} is denoted $p\times q$
and is given by 
\[
(p\times q).(s,s')=p.s\land q.s'
\]
The \emph{product} (parallel composition)
of two monotonic transformers $S$ and $T$ is
denoted by $S\times T$ and is given by 
\[
(S\times T).q.(s,s')=(\exists p,r:p\times r\le q\land S.p.s\land T.r.s')
\]
If ${\cal{S}}=\{S_{i}\}_{i\in I}$, is a collection of monotonic transformers,
then the \emph{fusion} of $\cal{S}$, denoted $\fusion.\cal{S}$, is a monotonic transformer
given by 
\[
\fusion.{\cal S}.q.s=\big(\exists p:(\bigwedge_{i\in I}p_{i})\le q\land(\forall i:S_{i}.p_{i}.s)\big)
\]
When $\cal S$ contains just two transformers $S$ and $T$, we write
$\fusion(S,T)$ instead of $\fusion(\{S,T\})$.

Note that product and fusion are similar, but not the same. 
The result of the product $A\times B$ is a component with separate inputs
and outputs for $A$ and $B$ (see Figure~\ref{fig_compos},
parallel composition), whereas the result of $\fusion(A,B)$
is a component with ``shared'' input and output.

\begin{thm}
\label{thm:product-fusion}Fusion and product satisfy the following~\cite{back:butler:1995}\end{thm}
\begin{enumerate}
\item $\fusion.( \{\{p_{i}\}\comp[r_{i}]\}_{i\in I})=\{\bigwedge_{i}p_{i\in I}\}\comp[\bigwedge_{i\in I}r_{i}]$ 
\item $S\sqsubseteq S'$ and $T\sqsubseteq T$' implies $S\times T\sqsubseteq S'\times T'$
\item $(\forall i:S_{i}\sqsubseteq T_{i})$ implies $\fusion.{\cal S}\sqsubseteq\fusion.{\cal T}$ 
\item $(\{p\}\comp[r])\times(\{p'\}\comp[r'])=\{x,y\;|\; p.x\land p'.y\}\comp[x,y\leadsto x',y'\;|\; r.x.x'\land r'.y.y']$ 
\end{enumerate}

\ifdefined\techrep
\subsection{Strictly conjunctive predicate transformers}

Dijkstra \cite{dijkstra-75} introduced predicate transformers for
reasoning about programs. In addition to monotonicity, he also required
some other properties, like \emph{conjunctivity} ($S.(p\land q)=S.p\land S.q$),
and the \emph{law of excluded miracle} ($S.\bot=\bot$) for his predicate
transformers. In this paper we use \emph{strictly conjunctive} predicate
transformers. These are the predicate transformers that satisfy a
conjunctivity property stronger than the one used by Dijkstra. A predicate
transformer is strictly conjunctive if it commutes with arbitrary
non-empty meets: 
\[
S.(\bigwedge_{i\in I}q_{i})=\bigwedge_{i\in I}S.q_{i}
\]
for all nonempty $I$ and $q:I\to Y\to\bool$.

The set of strictly conjunctive predicate transformers are exactly
the predicate transformers of the form $\assert p\comp\assume r$,
and the set of strictly conjunctive predicate transformers satisfying
the law of excluded miracle are exactly the predicate transformers
of the form $\assert{\inp.r}\comp\assume r$: 
\begin{thm}\cite{back-wright-98}
A predicate transformers $S$ is strictly conjunctive if and only if there
exist a predicate $p$ and a relation $r$ such that 
\[
S=\assert p\comp\assume r
\]
$S$ is strictly conjunctive and $S$ satisfies the law of
excluded miracle, if and only if there exists a relation $r$ such that 
\[
S=\assert{\inp.r}\comp\assume r.
\]
\end{thm}
For instance, for the predicate transformer $\skipstm$ we have
$$
\skipstm = \assert \top = \assume {x\leadsto x} = 
\assert \top \comp \assume {x\leadsto x}.
$$
For the predicate transformers
  $\assert p$, $\assume r$,
$\fail$ and $\magic$ we have:
$$
\begin{array}{lll}
\assert p = \assert p \comp \assume {x\leadsto x} &\ \ \ \ &
\fail = \assert \bot = \assert \bot \comp \assume {x\leadsto x}
\\[1ex]
\assume r = \assert \top \comp \assume r &&
\magic = \assume \bot = \assert \top \comp \assume \bot
\end{array}
$$
Therefore all these transformers are strictly conjunctive, but only $\skipstm$,
$\assert p$ and $\fail$ satisfy the law of excluded miracle. For example
we have $\assert p.\bot = p\land\bot=\bot$. On the other hand $\magic.\bot=\top$,
so $\magic$ does not satisfy the law of excluded miracle.

Consider also this set of examples:
$$
\begin{array}{lllllllll}
T_1 & = & \assume {x\leadsto y \;|\; y = x} = \assume {x\leadsto x} & \ \ \ \ 
   &T_4 & = & \assert{x\ge 0} \comp \assume {x\leadsto y \;|\; y = x} \\[1ex]
   
T_2 & = & \assume {x\leadsto y \;|\; x \ge 0 \land y = x} &&
   T_5 & = & \assert{x\ge 0} \comp \assume {x\leadsto y \;|\; x \ge 0 \land y = x}\\[1ex]
   
T_3 & = & \assume {x\leadsto y \;|\; x \ge 0 \impl y = x} &&
T_6 & = & \assert{x\ge 0} \comp \assume{x\leadsto y\;|\; x\ge 0 \impl y=x} 
\end{array}
$$
All these transformers are strictly conjunctive.
$T_1$, $T_2$, $T_3$ and $T_4$ are all different. $T_4$, $T_5$ and $T_6$ are all equal because the implication
$$(\forall x, y : p.x \impl r.x.y = r'.x.y) \impl
\assert p \comp \assume r = \assert p \comp \assume {r'}$$
is valid (i.e., if for any legal input $x$ the outputs of $r$ and $r'$ are
the same, then the transformers $\assert p \comp \assume r$ and
$\assert p \comp \assume r'$ are the same).
All inputs are legal for $T_1$, $T_2$, and $T_3$, but only inputs $x\ge0$ are legal for $T_4$.
Transformers $T_1$, $T_3$ and $T_4$ satisfy the law of excluded miracle because, for example
$\inp. (\lambda x,y: x \ge 0 \impl y = x) = \top$ and
$T_3 = 
   \assert {\inp. (\lambda x,y: x \ge 0 \impl y = x)} 
    \comp \assume {x\leadsto y \;|\; x \ge 0 \impl y = x}$.
$T_2$ does not satisfy the law of excluded miracle
because $T_2.\bot. x = (x < 0) \not=\fals$.
If we run all $T_1$, $T_2$, $T_3$ and $T_4$ on the same input $x \ge 0$, then they all 
return output $y = x$. However if we run them on the input $x < 0$, then they all behave diferently.
$T_1$ returns output $y = x$. 
$T_2$ behaves magically because the realation $x \ge 0 \land y = x$
is false. $T_3$ returns some arbitrary $y$ because the relation $x \ge 0 \impl y = x$
is true. $T_4$ fails for $x$ because $x$ is an illegal input for $T_4$.

For strictly conjunctive predicate transformers the
weak iteration is equal to an enumerable demonic choice: 
\begin{thm}
If $S$ is strictly conjunctive then 
\[
S^{*}=\bigsqcap_{n\in\nat}S^{n}
\]
\end{thm}

This theorem gives a simpler definition for the weak iteration $S^{*}$
when $S$ is strictly conjunctive, as compared to the case when $S$
is just monotonic. This theorem will be used later in Section
\ref{sub:rel-pt} to connect the definition of instantaneous feedback
on relations to the corresponding definition on predicate transformers.
\fi

\section{Relations with Fail and Unknown}

In this section we formally introduce relations with fail and unknown
values. As mentioned in the introduction, relations with fail and
unknown and strictly conjunctive predicate transformers have the same
expressive power, but each is better suited in defining, explaining,
or proving certain results.

\subsection{Relations with fail}

First we introduce relations augmented with a special element $\bullet$
which we call ``fail''. 
\ifdefined\techrep
These relations are equivalent to strictly
conjunctive predicate transformers \cite{back-wright-98}. If in addition
these relations are total, then they are equivalent to predicate transformers
that are strictly conjunctive and satisfy the law of excluded miracle
\cite{back-wright-98}.
\fi
Relations with fail
are used in the semantics of imperative programming languages to model
non-termination in the context of non-determinism \cite{plotkin:1976}.
In this paper $\bullet$ is used to model non-input-receptive systems:
when the output is $\bullet$, it means that the input is illegal.

We assume that $\bullet$ is a new element that does not belong to
ordinary types. We define the type $A^{\bullet}=A\cup\{\bullet\}$.
\ifdefined\techrep
An element of $x\in A^{\bullet}$ is either $\bullet$ or $x\in A$.
\fi
An element $x\in(A\times B)^{\bullet}$ is either $\bullet$ or it is a pair
$(a,b)$, with $a\in A$ and $b\in B$. A \emph{relation with fail} from
$A$ to $B$ is an element from $A^{\bullet}\to B^{\bullet}\to\bool$.
We assume that relations with fail 
map $\bullet$ only to $\bullet$, i.e., $R.\bullet.\bullet$ is true,
and $R.\bullet.x$ is false for all $x\not=\bullet$. If $R$ is a
relation with fail, and $R.x.\bullet$, then $x$ is \emph{an illegal
input} for $R$.

\ifdefined\techrep
To define a relation with fail $R$, it is enough to define $R.x.\bullet$
for $x\not=\bullet$, and to define $R.x.y$ for $x\not=\bullet$
and $y\not=\bullet$, and we assume also $R.\bullet.\bullet$. Similarly
when we state that a relation $R$ has certain properties, we state
these two cases, and in addition we also require $R.\bullet.\bullet$.

The condition that relations with fail map $\bullet$ to $\bullet$
is needed to ensure consistency of the serial composition of relations
with respect to illegal inputs. For example if $R$ and $R'$ are
relations with fail of appropriate types, and $x$ is an illegal input
for $R$, ($R.x.\bullet$ is true), then $x$ must also be an illegal
input for the serial composition $R\circ R'$. This is true because
$R.x.\bullet$ and $R'\bullet.\bullet$ implies $(R\circ R').x.\bullet$.
\fi

\begin{example}[Division]
\label{ex_division}
$\divrel:(\real^{2})^{\bullet}\to\real^{\bullet}\to\bool$
is a relation with fail modeling division:
\[
\divrel.(x,y).\bullet = (y=0) \mbox{ and } 
\divrel.(x,y).z = (y\not=0\land z=x/y).
\]
Because $\divrel.(x,0).\bullet$ is true, we have that $(x,0)$ is
an illegal input for $\divrel$, for any $x$.
\end{example}

If $x$ is an illegal input for $R$, then it is not important if
$R$ also maps $x$ to some proper output $y$ ($R.x.y=\tru$, with $y\neq\bullet$). If
$R'$ is a relation almost identical to $R$ except that $R'.x.y=\fals$,
then we consider $R$ and $R'$ as modeling the same system.
We do not restrict relations with fail to relations
that map illegal inputs only to $\bullet$ because we want to model
the {\em demonic choice} of relations by union. If $R$ and $R'$ are of the same type,
then $R\vee R'$ models the demonic choice of $R$ and $R'$.
If $x$ is illegal for $R$, then $x$ is illegal also
for $R\vee R'$, even if $x$ is legal for $R'$.

The relation $\rfail$ is defined by $(\forall x, y: \rfail.x.y = (y = \bullet))$. All inputs are illegal for $\rfail$. 
Similarly to predicate transformers we use the
syntax $(x \leadsto e)$ for the relation $R$ that satisfies $R.\bullet.\bullet$
and $(\forall x\not=\bullet: R.x.e)$, where $e$ is an expression containing $x$.

Relations with fail are closed under relation composition (serial
composition), and union (demonic choice).

\begin{defn}
For $R,R':A^{\bullet}\to B^\bullet\to\bool$ two relations with fail of the
same type, we define the \emph{refinement} $R\sqsubseteq R'$ by $(\forall x:R.x.\bullet\lor(R'.x\le R.x))$.
\end{defn}

\ifdefined\techrep
\subsection{Relations with fail and predicate transformers}
For a relation $R:A^{\bullet}\to B^{\bullet}\to\bool$ we introduce
its \emph{corresponding predicate transformer}. We define the \emph{precondition}
of $R$ ($\pre.R:A\to\bool$), the \emph{proper relation} of $R$
($\rel.R:A\to B\to\bool$), and the \emph{predicate transformer} (\emph{weakest
precondition}) of $R$ ($\wpr.R:(B\to\bool)\to(A\to\bool)$) by 
\[
\begin{array}{lll}
\pre.R.x & = & \neg R.x.\bullet\\
\rel.R.x.y & = & (R.x.y\land x\not=\bullet\land y\not=\bullet)\\
\wpr.R & = & \{\pre.R\}\circ[\rel.R]
\end{array}
\]

The predicate $\pre.R.x$ is true if $x$ is a legal input for $R$,
and $\rel.R.x.y$ is true if the system modeled by $R$ produces non-deterministically
the output $y$ from input $x$. $\wpr.R$ is the predicate transformer
corresponding to $R$, and it has the weakest precondition interpretation
as explained in Section \ref{sub:refinement}.

For the $\divrel$ relation we have $\pre.\divrel.(x,y)=(y\not=0)$,
and $\rel.\divrel.(x,y).z=(y\not=0\land z=x/y)$.
\fi

\ifdefined\techrep
Next lemma shows that $\wpr$ maps the operations on relations to
the corresponding operations on predicate transformers. 
\begin{lem}
\label{lem:wp:commute}The following hold:\end{lem}
\begin{enumerate}
\item $\pre.(R\vee R')=\pre.R\wedge\pre.R'$ 
\item $\pre.(R\circ R').x=\pre.R.x\land(\forall y.\rel.R.x.y\impl\pre.R'.y)$ 
\item $\rel.(R\vee R')=\rel.R\vee\rel.R'$ and $\rel.(R\circ R')=\rel.R\circ\rel.R'$ 
\item $\wpr.(R\vee R')=\wpr.R\sqcap\wpr.R'$ and $\wpr.(R\circ R')=\wpr.R\comp\wpr.R'$ 
\end{enumerate}
If we replace the binary unions (respectively, intersections) in the
above theorem by arbitrary non-empty unions (resp., intersections),
then the results still hold.
\fi

\ifdefined\techrep
For a relation $R:A\to A\to\bool$ the reflexive and transitive closure
of $R$ and the transitive closure of $R$, denoted $R^{*}$ and $R^{+}$,
are relations from $A\to A\to\bool$ defined by 
\[
R^{*}=\bigvee_{n:\nat}R^{n}\ \ \ R^{+}=\bigvee_{0<n}R^{n}
\]
where $R^{n}$ is $R$ composed with itself $n$ times ($R\circ R\circ\cdots\circ R$).

If $R$ is a relation with fail, then $R^{*}$ is also, and the following
lemma holds 
\begin{lem}
\label{lem:star}If $R:A^{\bullet}\to A^{\bullet}\to\bool$ is a relation
with fail then 
\[
\wpr.(R^{*})=\bigsqcap_{n:\nat}(\wpr.R)^{n}=(\wpr.R)^{*}
\]
\end{lem}

Next lemma shows that the refinement of relations is preserved by
the operations $\vee$, $\circ$, and $*$. 
\begin{lem}
\label{lem:relation:refinement}Let $R,R',U,U'$ be relations with fail
of appropriate types, then\end{lem}
\begin{enumerate}
\item $R\sqsubseteq R'\Leftrightarrow \wpr.R\sqsubseteq\wpr.R'$ 
\item $R\sqsubseteq R'\land U\sqsubseteq U'\impl R\circ U\sqsubseteq R'\circ U'$
and $R\vee U\sqsubseteq R'\vee U'$ 
\item $R\sqsubseteq R'\impl R^{*}\sqsubseteq R'^{*}$.
\end{enumerate}
\fi

\subsection{Unknown values}

We denote by $\bot$ a special element called {\em unknown} which is not
part of ordinary types: $\nat$, $\bool$, $\real$, .... For
a set $A$ we define a partial order on $A_{\bot}=A\cup\{\bot\}$
so that, for $a,b\in A_{\bot}$:
\[
a\le b \;\; \Leftrightarrow \;\; a=\bot\lor a=b
\]
This order is the \emph{flat complete partial order} (flat CPO) \cite{priestley}.

We extend two partial orders on $A$ and $B$ to a partial order on
the Cartesian product $A\times B$ by 
\[
(a,b)\le(a',b')\Leftrightarrow a\le a'\land b\le b'
\]
$\bot$ is the smallest element of $A_{\bot}$,
and for the Cartesian product $A=A_{1\bot}\times\ldots\times A_{n\bot}$we
denote also by $\bot$ the tuple $(\bot,\ldots,\bot)\in A$. 
\ifdefined\techrep
The element
$\bot$ of $A$ is the smallest element of $A$ with the order extended
to Cartesian product.
\fi
For simplicity, we call the Cartesian product of a collection of flat CPOs also
a flat CPO.

For partial order $(A,\le)$, an element $m\in A$ is \emph{maximal}
if there is no element in $A$ greater than $m$: 
\[
\maximal.m=(\forall x\in A:m\not\le x)
\]

The context will distinguish the cases when $\bot$ is an element
of $A_{\bot}$, or the smallest predicate or relation. However, in
all cases $\bot$ denotes the smallest element of some partially ordered
set.

\begin{example}[AND gate with unknown]
Consider the function
 $\land_{\bot}:\bool_{\bot}\to\bool_{\bot}\to\bool_{\bot}$
defined in Table \ref{tab:and-gate}. It models a logical AND gate
which returns false if at least one of its inputs is false,
even when the other input is unknown.
We will return to this example in Example~\ref{ex_AND_feedback}.
\end{example}
\begin{table}
\begin{centering}
\begin{tabular}{|c|c|c|c|}
\hline 
$\land_{\bot}$  & $\bot$  & $\fals$  & $\tru$\tabularnewline
\hline 
$\bot$  & $\bot$  & $\fals$  & $\bot$\tabularnewline
\hline 
$\fals$  & $\fals$  & $\fals$  & $\fals$\tabularnewline
\hline 
$\tru$  & $\bot$  & $\fals$  & $\tru$\tabularnewline
\hline 
\end{tabular}
\par\end{centering}

\centering{}\protect\protect\caption{AND gate with unknown\label{tab:and-gate}}
\end{table}

\subsection{Relations with fail and unknown (RFUs)}

We use unknown values ($\bot$) as the starting value of the feedback
variable in an iterative computation of the feedback, and we use $\bullet$
to represent ``failure'' (e.g. illegal inputs). To see why we distinguish
the two, suppose we used the same value (say $\bot$) to represent
both these concepts. Suppose $R.1.\bot$ is true for some relation
$R$. This could mean that $1$ is an illegal input for $R$, or that
given $1$ as input, the output is unknown. Now, suppose we have another
relation $R'$ that maps unknown inputs to $0$, and we compose $R$
and $R'$ in series. What should the composition $R\circ R'$ produce
given input $1$? Is $1$ illegal for $R\circ R'$, or does $R\circ R'$
map $1$ to $0$, or both? To avoid such ambiguities, we separate
the two concepts and values.

\begin{figure}
\centering
		\includegraphics{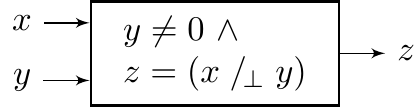}
		\includegraphics{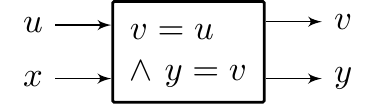}
\protect\protect\caption{\label{figRFUs} Examples of RFUs: $\divrel$ (left),
and $\mathsf{ID}$ (right)}
\end{figure}

\begin{example}[Division with fail and unknown]
Continuing Example~\ref{ex_division}, we extend relation $\divrel$
with unknown.
If the input $y$ is zero, then $\divrel.(x,y)$ fails, otherwise
if $x$ is zero, then the result $z$ is zero, even if $y$ is unknown.
The complete definition of $\divrel$ is: 
\[
\begin{array}{ll}
\mathsf{\divrel}.(x,y).\bullet \;\;=\;\;  (y=0)\\
\mathsf{\divrel}.(x,y).z \;\;=\;\;  (y\not=0)\land\big((x=0\land z=0)
\ifdefined\techrep \lor \else\\ \ \ \  \ \ \lor\ \fi 
  (x\not=0\land y=\bot\land z=\bot)
 \lor (x=\bot\land z=\bot)
\ifdefined\techrep \\\hspace{10cm}\lor \ \else\\ \ \ \  \ \lor\ \fi
(x\not=\bot\land y\not=\bot\land z=x/y)\big).
\end{array}
\]
$\divrel$ is illustrated in Figure \ref{figRFUs}, where
we use $/_{\!\bot}$ for the division operator $/$ 
extended to unknown values as defined above.
\end{example}

\section{\label{sec:inst-feedback}Instantaneous Feedback}

In this section we introduce the definition of instantaneous feedback
for an RFU
$R:(A\times B)^{\bullet}\to(A\times C)^{\bullet}\to\bool$,
as shown in Figure \ref{fig:instant}(a), where 
$A=A_{1\bot}\times\ldots\times A_{n\bot}$. 
The intention is to feed output $v$ back into input $u$, as shown
in Figure \ref{fig:instant}(b).
The fact that we require $R$ to have an extra input $x$ and an extra output $y$
is without loss of generality, as we can model absence of these wires by having
their type ($B$ or $C$) be the {\em unit type}, i.e., the type that has 
the empty tuple () as its only element.

\begin{figure}[h]
\centering{}\includegraphics{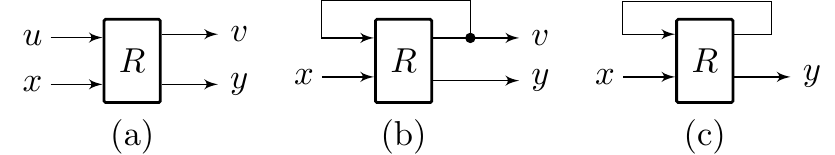}\protect\protect\caption{(a) RFU $R$, (b) $\protect\fbfun.R$, (c) $\protect\fb.R$\label{fig:instant}}
\end{figure}

The idea of the instantaneous feedback calculation is outlined next.
We start with $\bot=(\bot,\ldots,\bot)\in A$, and a given $x\in B$,
and we construct the following tree: The nodes of the tree are labeled
with elements in $(A\times (C \cup\{?\}))^{\bullet}$.
The root of the tree is $(\bot,?)$, and for every node of the tree of 
the form $(u,?)$ we create new children nodes as follows:
\begin{enumerate}
\item If $R.(u,x).\bullet$ then we create the child $\bullet$.
\label{treerule1}
\item If there exists $v, y$ such that $u < v $ and $ R.(u,x).(v, y)$, then
  we create the child $(v,?)$.
\label{treerule2}
\item If there exists $y$ such that $R.(u,x).(u, y)$, then
  we create the child node $(u,y)$.
\label{treerule3}
\end{enumerate}

We thus obtain a tree with leaves of the form $\bullet$,
$(u,?)$, or $(u,y)$.
Note that the depth of this tree is finite, because whenever we create a
child $(v, ?)$ of $(u, ?)$ we require $u < v$,
and we cannot have a strictly increasing
infinite sequence of $A$ elements (because $A$ is a product of flat CPOs).
That is, if all $u$ values become known (non-$\bot$), then all children
of $(u,?)$ are leaves.

Let us illustrate this tree construction with some  examples.
\begin{example}[The trees of the relations $\tru$ and $u\neq v$ discussed in Section~\ref{sec_intro}]
\label{ex_trees}
Consider the RFUs $\mathsf{TRUE}$ and $\mathsf{NEQ}$, having inputs $u,x\in\{0,1\}_\bot$ and outputs $v,y \in \{0,1\}_\bot$, defined by:
\[
\begin{array}{l}
  \mathsf{TRUE}.(u,x).(v,y) \;\; = \;\; (v=y)\\[1ex]
\mathsf{NEQ}.(u,x).(v,y) \;\; = \;\; (v=y)\land (u=\bot\impl v=\bot)
\ifdefined\techrep \land \else \\ \qquad\qquad\qquad\qquad\qquad\land\ \fi
  (u\not=\bot\impl (v\not=u\land v\not=\bot)).
\end{array}
\]

$\mathsf{TRUE}$ is the RFU
corresponding to the relation $\tru$ discussed in the introduction 
(this system is also sometimes called {\em havoc}). 
$\mathsf{TRUE}$ accepts any input (i.e., is input-receptive)
and non-determini\-stically returns any output.
Note that its two outputs $v$ and $y$ can be anything,
but are always equal due to the constraint $v=y$.
$\mathsf{NEQ}$ is the RFU corresponding to the relation $u\neq v$ 
discussed in the introduction (component $A'$).
$\mathsf{NEQ}$ accepts any input $x$ and returns any output $v$
different from that input (again, $y$ is always equal to $v$).
Since $\mathsf{NEQ}$ is an RFU, we also define how the relation 
behaves with unknown values. 
\ifdefined\techrep
$\mathsf{TRUE}$ and $\mathsf{NEQ}$ are illustrated in
Figure~\ref{fig:havoc} and Figure~\ref{fig:diff}, respectively.
\begin{figure}[h]
\centering{}
	\subfloat[]{\label{fig:havoc}
		\includegraphics{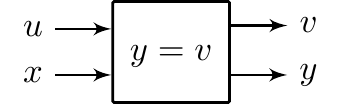}}	
	\subfloat[]{\label{fig:diff}
		\includegraphics{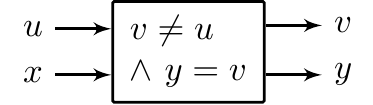}}	
\protect\protect\caption{(a) $\mathsf{TRUE}$, (b) $\mathsf{NEQ}$}
\end{figure}
\fi

The trees of $\mathsf{TRUE}$ and $\mathsf{NEQ}$
are shown in Figure~\ref{fig:treea}.
Both trees are independent of the value of input $x$.
In the tree of $\mathsf{TRUE}$ we start with $u=\bot$ and according to
the tree construction rule~\ref{treerule2} we create the children with
$v=0$ and $v=1$, since $v>\bot$ in both these cases.
According to rule~\ref{treerule3}, we also create the child $(\bot,\bot)$,
because $\mathsf{TRUE}.(\bot,x).(\bot,\bot)$ is true for any $x$.
From node $(0,?)$ the only possible child is $(0,0)$, according to
rule~\ref{treerule3},
because $\mathsf{TRUE}.(0,x).(0,0)$ is true for any $x$.
Note that, even though $\mathsf{TRUE}$ allows $v$ to be arbitrary
when $u=0$, for all such $v$, the ordering $u<v$ does not hold due to the
flat CPO property. Therefore, $(0,0)$ is the only possible child of $(0,?)$.
Similarly for $(1,?)$ and $(1,1)$.
In the tree of $\mathsf{NEQ}$, when starting with $u=\bot$, we can only choose 
$v=\bot$. 
\end{example}
\begin{figure}[h]
\ifdefined\techrep
\centerline{\includegraphics{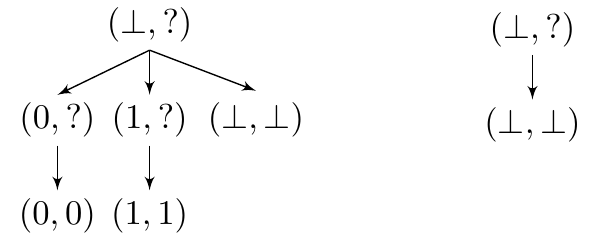}}
\else
\centerline{\includegraphics[width=.7\columnwidth]{fig/treea}}
\fi
\caption{Tree for $\mathsf{TRUE}$ (left); Tree for $\mathsf{NEQ}$ (right)
\label{fig:treea}}
\end{figure}
Given the above tree construction, the \emph{instantaneous feedback} of an RFU
$R$, denoted $\fbfun.R$, will be a relation from $B^{\bullet}$ to
$(A\times C)^{\bullet}$ such that: $\fbfun.R.x.(u,y)$ is true if
$(u,y)$ is a leaf in the tree; and $\fbfun.R.x.\bullet$ is true
if $\bullet$ is a leaf in the tree. 
The type of $\fbfun.R$
is shown in Figure \ref{fig:instant}(b).

Next we define $\fbfun.R$ formally.
In addition we define another
version of the feedback ($\fb.R:B^{\bullet}\to C^{\bullet}$) in which
we remove the output component $u$ if $u$ is maximal (i.e., if it
contains no unknowns), or we make the feedback fail otherwise. The
type of $\fb.R$ is shown in Figure~\ref{fig:instant}(c). 
\begin{defn}
\label{def:inst-feedback}For an RFU
$R:(A\times B)^{\bullet}\to(A\times C)^{\bullet}\to\bool$
we define $\fbfun.R:B^{\bullet}\to(A\times C)^{\bullet}\to\bool$,
$\fbend:(A\times C)^{\bullet}\to C^{\bullet}\to\bool$, and $\fb.R:B^{\bullet}\to C^{\bullet}$:
\[
\begin{array}{l}
\fbfun.R.x.(u,y)\\
\quad=\exists n,u_{0},\ldots,u_{n}:u_{0}=\bot\land u_{n}=u\land R.(u,x).(u,y)\\
\qquad\land\ \big(\forall i<n:u_{i}<u_{i+1}\land(\exists z:R.(u_{i},x).(u_{i+1},z))\big)\\
\fbfun.R.x.\bullet\\
\quad=\exists n,u_{0},\ldots,u_{n}:u_{0}=\bot\land R.(u_{n},x).\bullet\\
\qquad\land\ \big(\forall i<n:u_{i}<u_{i+1}\land(\exists z:R.(u_{i},x).(u_{i+1},z))\big)\\
\fbend.(u,y).z=\maximal.u\land z=y\\
\fbend.(u,y).\bullet=\neg\maximal.u\\
\fb.R=\fbfun.R\circ\fbend
\end{array}
\]
\end{defn}

\begin{example}[Instantaneous feedback of $\mathsf{TRUE}$ and $\mathsf{NEQ}$]
\label{ex_TRUE_NEQ}
Continuing Example~\ref{ex_trees}, let us compute the instantaneous feedback
operators defined above on the RFUs $\mathsf{TRUE}$ and $\mathsf{NEQ}$.
We get:
\[
\begin{array}{l}
\fbfun.\mathsf{TRUE}.x.(v,y)\;\;=\;\; (v=y)\\[1ex]
\fbfun.\mathsf{NEQ}.x.(v,y)\;\;=\;\; (v=y=\bot)\\[1ex]
\fb.\mathsf{TRUE}\;\;=\;\; \rfail\\[1ex]
\fb.\mathsf{NEQ}\;\;=\;\; \rfail
\end{array}
\]
As we should expect, the result of our instantaneous feedback definition is $\rfail$, for both relations.
This is very different from the straightforward definition discussed
in the introduction, and eliminates the problems presented there.
Specifically, the refinement $\mathsf{TRUE}\isrefinedby\mathsf{NEQ}$
is preserved by feedback, solving the problem identified
in the introduction.
The result of applying $\fb$ to $\mathsf{TRUE}$ and $\mathsf{NEQ}$
is $\rfail$ because as can be seen from the trees of
Figure~\ref{fig:treea}, the feedback variable $u$ can be assigned any value, 
including the unknown value $\bot$. 
\end{example}

We next provide more examples that 
illustrate our instantaneous feedback operator.
We also prove an important result which shows that the operator
reduces to standard operators for the special case of deterministic
and input-receptive systems.

\begin{figure}[h]
\centering{}\includegraphics{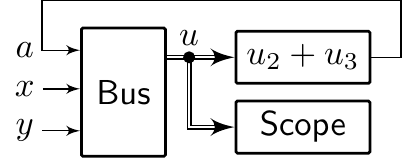}\protect\protect\caption{Bus followed by function block in feedback\label{fig_bus}}
\end{figure}

\begin{example}[A seemingly algebraic loop in a Simulink diagram]
\label{ex_simulink}
The diagram shown in Figure~\ref{fig_bus} is inspired from 
real-world Simulink models. This particular pattern is found in
an automotive fuel control system benchmark by Toyota~\cite{jin2014ARCH,JinDKUB14}.
In the picture, $a$, $x$, and $y$ are inputs to a {\em Bus} block which outputs
them as a tuple. This tuple is then fed into a function block which performs
addition on the 2nd and 3rd elements of the tuple ($u_2$ and $u_3$).
The output of the function is fed back into the bus. 
But there is no algebraic loop because the function does not depend on
the 1st element of the tuple, i.e., $a$.
The output of the bus is fed also into a {\em Scope} block which is supposed
to print all three elements of the tuple on the screen. We can model the bus and
the scope as the identity function, and we have:
$$
  \begin{array}{lll}
  \mathsf{Bus}.(a,x,y).(u_1,u_2,u_3) = (a = u_1 \land x = u_2 \land y = u_3)\\
  \mathsf{Scope}.(u_1,u_2,u_3).(a,x,y) = (a = u_1 \land x = u_2 \land y = u_3)\\
  \mathsf{Fun}.(u_1,u_2,u_3).v = (v = u_2 + u_3)\\
  \mathsf{Sys} = \mathsf{Bus} \circ (u \leadsto u,u) \circ (\mathsf{Fun}\times \mathsf{Scope})
  \end{array}
$$
where $\mathsf{Sys}$ is the system without the feedback connection. If we
apply the feedback we obtain, as could be expected:
$$
  \fbfun.\mathsf{Sys} = (x,y \leadsto x+y,x,y)
$$
\end{example}
\begin{figure}
\centering{}\includegraphics{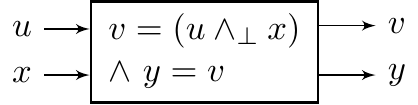}\protect\protect\caption{AND gate\label{fig_andgate}}
\end{figure}
\begin{example}[Instantaneous feedback of AND gate]
\label{ex_AND_feedback}
Suppose we connect in feedback the AND gate function
from Table \ref{tab:and-gate}. To apply feedback, we again add an extra
output $y=v$ (see Figure~\ref{fig_andgate}), and we get:
\[
\begin{array}{l}
\mathsf{AND}.(u,x).(v,y)\;\;=\;\; (v=y=u\,\land_{\bot}\, x)\\[1ex]
\fbfun.\mathsf{AND}.x.(v,y)\;\;=\;\; 
\ifdefined\techrep\else\\\quad\fi
(v=y)
\land (x=\fals\impl y=\fals)\land(x\not=\fals\impl y=\bot)\\[1ex]
\fb.\mathsf{AND}.x.y \;\;=\;\; (x = \fals \impl y = \fals)
\ifdefined\techrep\land\else\\ \quad \land \ \fi
(x=\tru \impl y = \bullet).
\end{array}
\]
The result of the feedback is a component which outputs
false when its free input $x$ is false. On the other hand, if $x$ is
true or unknown, then the component fails.
\end{example}

The result above is consistent with the result obtained using classic
theories of instantaneous feedback for total functions (or
{\em constructive semantics})~\cite{Malik94,Berry99,EdwardsLee03}.
This will generally be the case for deterministic systems, as the tree
described in the instantaneous feedback calculation above, becomes a single
path in those cases. Therefore, for the special case of total functions,
our semantics specializes to the constructive semantics.

Let us state this formally. 
Let $A$ be a flat CPO, and let 
$R:(A\times B)^{\bullet}\to(A\times C)^{\bullet}\to\bool$ be a relation
such that there exist total functions $f:B\to A \to A$ and $g:B\to A\to C$
such that for all $x$, $f.x$ is monotonic on $A$ 
and $\forall u,x,z: R.(u,x).z \Leftrightarrow z = (f.x.u,g.x.u)$. 
Then:
\begin{thm}
If $A$ is a flat CPO and $R$ is as above then
$$\fbfun.R.x.(u,y) \iff u = \mu\, (f.x) \land y = g.x.u$$
where $\mu\, (f.x)$ is the least fixpoint of $f.x$, which exists because
of the assumption that $f.x$ is monotonic on $A$. 
\end{thm}

\ifdefined\techrep
\begin{example}[AND gate with an additional wire]
\[
\begin{array}{l}
\mathsf{AND}'.((w,u),x).((v,w'),y) \;\;=\;\; (w=w'\land v=y=u\,\land_{\bot}\, x)\\
\fbfun.\mathsf{AND}'.x.((v,w'),y) \;\;=\;\; 
\ifdefined\techrep\else\\ \qquad\fi
(v=w'=y \land\ x=\fals\impl y=\fals \land x\not=\fals\impl y=\bot)\\
\fb.\mathsf{AND'}.x.y \;\;=\;\; (x = \fals \impl y = \fals)
 \land \ (x=\tru \impl y = \bullet).
\end{array}
\]
The feedback of this component is the same as the feedback for the
$\mathsf{AND}$ component. 
\end{example}

\begin{figure}
\centering{}\includegraphics{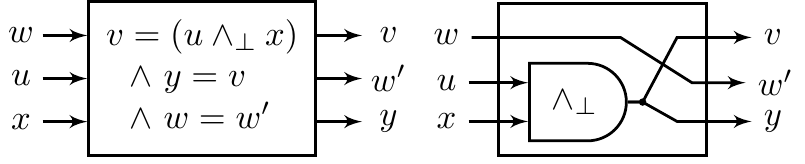}\protect\protect\caption{AND gate with an additional wire}
\end{figure}
\else
\fi

\begin{example}[Instantaneous feedback of $\mathsf{ID}$]
\label{ex_ID}
One might wonder what might happen if we connect the identity function
in feedback. Since our feedback operator takes relations with two input 
and two output wires, we define the identity operator
\[
\mathsf{ID}.(u,x).(v,y)\;\;=\;\; (u=v=y)
\]
which ignores input $x$, and assigns both outputs $v$ and $y$ to its
other input $u$ (see Figure \ref{figRFUs}).
Output $y$ allows to record the result of the feedback when we hide 
the feedback variables ($u$ and $v$). As might be expected, the result
of applying feedback is $\rfail$:
\[
\begin{array}{l}
\fbfun.\mathsf{ID}.x.(v,y)\;\;=\;\; (v=y=\bot)\\[1ex]
\fb.\mathsf{ID}\;\;=\;\; \rfail
\end{array}
\]
The result is $\rfail$ because the feedback variable is assigned the
value unknown. Again, this is consistent with the result obtained using
constructive semantics.
\end{example}

The power of our framework is that it applies not just to deterministic
and input-receptive systems, as in Examples~\ref{ex_simulink}, 
\ref{ex_AND_feedback}, and~\ref{ex_ID}, or to non-deterministic but still
input-receptive systems, as in Examples~\ref{ex_trees} and~\ref{ex_TRUE_NEQ},
but also to non-deterministic and non-input-receptive systems,
as in the example below.

\begin{example}[Instantaneous feedback of a non-deterministic and non-input-receptive system]
Consider the RFU $\mathsf{NonDet}$, having inputs $u\in\nat_\bot$ and $x\in\nat$ and outputs $v \in \nat_\bot$ and $y\in\nat$.
$\mathsf{NonDet}$ is defined as:
\[
\begin{array}{l}
\mathsf{NonDet}.(u,x).\bullet \;\;=\;\; (u = 2) \\
\mathsf{NonDet}.(u,x).(v,y) \;\; = \;\; (u \not=2) \land 
      (y = v \lor y = v +1) \ \land \\
      \ifdefined\techrep \hspace{6cm}\else\quad\fi \big(v = x + 1 \lor v = x + 2  
      \ifdefined\techrep \lor \else \ \lor \\ \qquad \fi 
       (u=\bot \land v = 6) \lor
     (u\not=\bot \land v=7)\big).
\end{array}
\]
That is, all inputs where $u = 2$ are illegal, and if
$u\not = 2$ then the output $v$ is non-deterministically $x+1$, or $x+2$,
or $6$ if $u=\bot$, or $7$ if $u\not=\bot$.
Note that if $u = \bot$ then $u \not=2$.
Output $y$ is either $v$ or $v + 1$.
\ifdefined\techrep
For $x=0$ and $x = 2$, we get the trees of Figure~\ref{fig:tree}.
\begin{figure}[h]
\centering{}\includegraphics{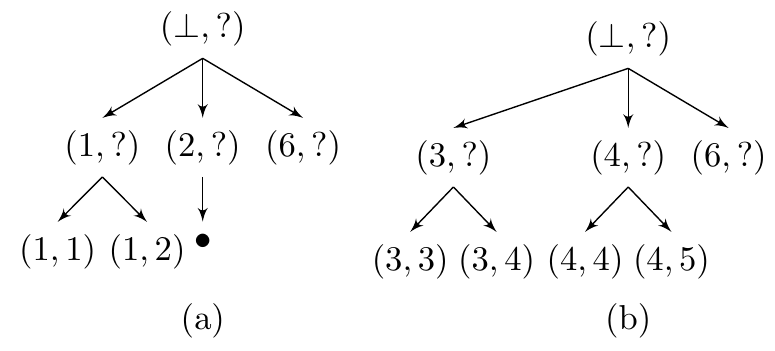}\protect\protect\caption{Trees for 
$\mathsf{NonDet}$ (a) $x=0$, (b) $x=2$
\label{fig:tree}}
\end{figure}

\fi
Then, we have: 
$$
\begin{array}{l}
\fbfun.\mathsf{NonDet}.x.\bullet \ \ = \ \ (x = 0 \lor x = 1)\\
\fbfun.\mathsf{NonDet}.x.(v,y) \ \ = \ \ \ifdefined\techrep\else\\\qquad\fi
  (x \not= 1 \land v = x +1 \land (y = x + 1 \lor y = x + 2)) \ \lor\\
  \ifdefined\techrep\hspace{5.25cm}\else\qquad\fi
	 (x \not= 0 \land v = x + 2 \land (y = x + 2 \lor y = x + 3))
\end{array}
$$
and because $v$ has always a known value after feedback, $\fb$ is the same, but without the 
output component $v$. We can also observe that in the feedback of $\mathsf{NonDet}$ 
the cases $v=6$ and $v=7$ do not occur. This is because the feedback does not stabilize
in these cases. 
\end{example}

One of the main results of this paper is that the feedback operation
preserves refinement: 
\begin{thm}
\label{thm:feedback:refinement}If $R,R':(A_{\bot}\times B)^{\bullet}\to(A_{\bot}\times C)^{\bullet}\to\bool$,
then 
\[
R\sqsubseteq R'\impl\fbfun.R\sqsubseteq\fbfun.R'.
\]
\end{thm}

\ifdefined\techrep
To prove this theorem we introduce some additional functions on $R$
and we use them to give an alternative point free definition of $\fb.R$.
\begin{figure}[h]
\centering{}\includegraphics{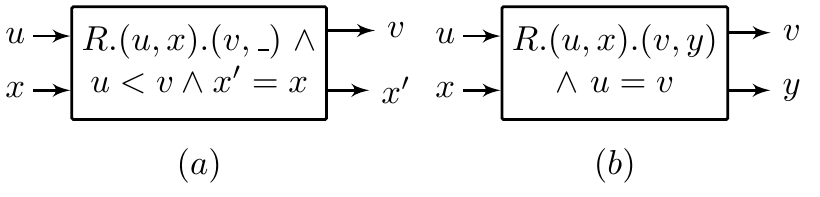}\protect\protect\caption{$(a)$ $\protect\fba.R$, $(b)$ $\protect\fbb.R$}
\end{figure}

\begin{defn}
For a relation $R$ from $(A_{\bot}\times B)^{\bullet}\to(A_{\bot}\times C)^{\bullet}\to\bool$
we define $\fbbegin:B^{\bullet}\to(A_{\bot}\times B)^{\bullet}\to\bool$,
$\fba:(A_{\bot}\times B)^{\bullet}\to(A_{\bot}\times B)^{\bullet}\to\bool$,
and $\fbb:(A_{\bot}\times B)^{\bullet}\to(A_{\bot}\times C)^{\bullet}\to\bool$
by 
\[
\begin{array}{l}
\fbbegin.x.(u,x')=(u=\bot)\land(x=x')\\
\fba.R.(u,x).(v,x') \ifdefined\techrep \else \\ \quad \fi
  =(u<v)\land(x=x')\land(\exists y:R.(u,x).(v,y))\\
\fba.R.(u,x).\bullet=R.(u,x).\bullet\\
\fbb.R.(u,x).(v,y)=(u=v)\land R.(u,x).(u,y)\\
\fbb.R.(u,x).\bullet=R.(u,x).\bullet
\end{array}
\]
\end{defn}

Next lemma shows that $\fba$ and $\fbb$ preserve refinement, and
$\fbfun.R$ can be defined as a point free combination of $\fbbegin$,
$\fba.R$, and $\fbb.R$: 
\begin{lem}
\label{lem:feedback:point:free}If $R,R':(A_{\bot}\times B)^{\bullet}\to(A_{\bot}\times C)^{\bullet}\to\bool$,
then\end{lem}
\begin{enumerate}
\item $\fbbegin$, $\fba.R$, $\fbb.R$, and $\fbend$ are relations with
fail. 
\item $R\sqsubseteq R'\impl\fba.R\sqsubseteq\fba.R'$ 
\item $R\sqsubseteq R'\impl\fbb.R\sqsubseteq\fbb.R'$ 
\item $\fbfun.R=\fbbegin\circ(\fba.R)^{*}\circ\fbb.R$ 
\end{enumerate}
Using this lemma and Lemma \ref{lem:relation:refinement} we can prove
now Theorem \ref{thm:feedback:refinement}:
Assume $R\sqsubseteq R'$, then we have 
\begin{lyxlist}{00.00}
\item [{~}] $\fbfun.R$ 
\item [{$=$}] \{Lemma \ref{lem:feedback:point:free}\} 
\item [{~}] $\fbbegin\circ(\fba.R)^{*}\circ\fbb.R$ 
\item [{$\sqsubseteq$}] \{Lemma \ref{lem:feedback:point:free} and Lemma
\ref{lem:relation:refinement}\} 
\item [{~}] $\fbbegin\circ(\fba.R')^{*}\circ\fbb.R'$ 
\item [{$=$}] \{Lemma \ref{lem:feedback:point:free}\} 
\item [{~}] $\fbfun.R'$\end{lyxlist}
\else
\fi

Another main result of this paper is that serial composition can be
expressed as parallel composition followed by feedback.
To state this, we first define the {\em cross product} of two relations,
which is essentially their parallel composition, but with their outputs
swapped, as shown in Figure \ref{fig:feedback:cross}a.
The swapping is done so that we can apply feedback to the right ``wires'',
as shown in Figure \ref{fig:feedback:cross}b.

\begin{figure}[h]
\centering{}\includegraphics{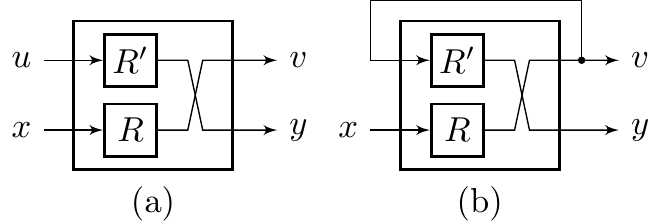}\protect\protect\caption{\label{fig:feedback:cross} (a) cross product, (b) feedback of cross
product}
\end{figure}
\begin{defn}
The cross product of two relations
with fail $R:A^{\bullet}\to B^{\bullet}\to\bool$ and $R:C^{\bullet}\to D^{\bullet}\to\bool$
is a relation with fail $R\otimes R':(C\times A)^{\bullet}\to(B\times D)^{\bullet}\to\bool$,
given by 
\[
\begin{array}{lll}
R\otimes R'.(u,x).\bullet & = & R.x.\bullet\lor R'.u.\bullet\\
R\otimes R'.(u,x).(v,y) & = & R.x.v\land R'.u.y
\end{array}
\]
\end{defn}

We can now prove that the feedback of the cross product of two
relations is the same as their serial composition:

\begin{thm}\label{thm:cross}
If $R:A_{\bot}^{\bullet}\to B_{\bot}^{\bullet}\to\bool$, $R':B_{\bot}^{\bullet}\to C_{\bot}^{\bullet}\to\bool$,
$R'$ is total and $\neg R'.\bot.\bullet$ then 
\[
\begin{array}{lll}
\fbfun.(R\otimes R').x.\bullet & =(R\circ R').x.\bullet\\
\fbfun.(R\otimes R').x.(u,y)\!\!\!\!\! & =R.x.u\land R'.u.y\\
\fb.(R\otimes R').x.\bullet & =(R\circ R').x.\bullet\lor R.x.\bot\\
\fb.(R\otimes R').x.y & =(\exists u:u\not=\bot\land R.x.u\land R'.u.y)
\end{array}
\]
\end{thm}
The condition $\neg R'.\bot.\bullet$ is needed because if $R'$ maps
unknown ($\bot$) to fail ($\bullet$), then the feedback iteration
stops and the whole left term maps some external input $x$ to fail.
On the other hand, if $R$ maps $x$ only to $a\not\in\{\bot,\bullet\}$,
and $R'$ maps $a$ only to $b\not=\bullet$, then, the serial composition
does not fail for the same $x$. This assumption is reasonable, since
all it states is that the unknown value must be legal, at least in
contexts when we want to use a component in feedback.

\ifdefined\techrep
\subsection{Associativity of instantaneous feedback}

In this section we show how to compute the feedback of a relation
$R:((A_\bot \times B_\bot) \times C)^\bullet \to ((A_\bot \times B_\bot)  \times D)^\bullet \to\bool$
as in Figure \ref{fig:compose-inst}(a) in two different
ways. First we compute the feedback on $(u,s)$-$(v,t)$, as in Figure \ref{fig:compose-inst}(b).
Second, we compute the feedback on $u$-$v$, and then on $s$-$t$,
as in Figure \ref{fig:compose-inst}(c). We would like that these two
composition patterns result in the same relation.
Indeed, we shall show that the relations of Figure \ref{fig:compose-inst}(b) and (c) are equivalent, under certain assumptions.

\begin{figure}[h]
\centering{}\includegraphics{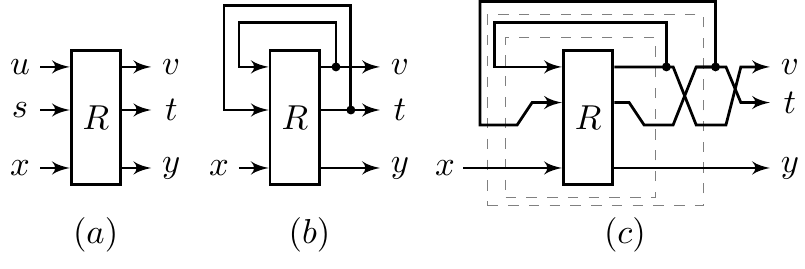}\protect\protect\caption{\label{fig:compose-inst}Composing instantaneous feedback}
\end{figure}

The feedback from Figure \ref{fig:compose-inst}(b) corresponds to $\fbfun.R$. 
Note that one application of $\fbfun$ is enough, because $R$ takes
as input $(u,s)$ as a pair, and outputs $(v,t)$ also as a pair.

To define the feedback from Figure \ref{fig:compose-inst}(c)
we need to introduce three adaptation functions: 
\[
\begin{array}{l}
\adapta.(u,(s,x)).((u',s'),x')=(u=u')\land(s=s')\land(x=x')\\
\adaptb.(v,(t,y)).(t',(v',y'))=(v=v')\land(t=t')\land(y=y')\\
\adaptc.(t,(v,y)).((v',t'),y')=(v=v')\land(t=t')\land(y=y')
\end{array}
\]
The composition $A=\adapta\circ R\circ\adapta^{-1}$ is represented
in the inner dashed rectangle from Figure \ref{fig:compose-inst}(c),
and it has the appropriate type to construct the feedback on the variables
$u$-$v$. The system $B=\fbfun.A$ has input $(s,x)$ and output
$(v,(t,y))$, and we want now to connect in feedback $s$ and $t$.
For this we need the second adapt function that switches $v$ and
$t$: $C=(B\circ\adaptb)$. System $C$ corresponds to the outer dashed
rectangle from Figure \ref{fig:compose-inst}(c). The system $\fbfun.C$
has input $x$ and output $(t,(v,y))$, and to this we apply $\adaptc$
to obtain $((v,t),y)$. We define $\seqfb.R=(\fbfun.C)\circ\adaptc$,
and we have: 
\[
\begin{array}{l}
\seqfb.R=(\fbfun.
((\fbfun.(\adapta\circ R\circ\adapta^{-1}))\circ\adaptb))\circ\adaptc
\end{array}
\]

In order to prove the equality $\fbfun.R=\seqfb.R$ we introduce some
additional assumptions. For a relation $R$ of appropriate type we
define:
\[
\begin{array}{l}
\monorel.R=
\quad(\forall u,u',x,v,y:u\le u'\land R.(u',x).(v,y)\impl R.(u,x).(v,y))\\[1ex]
\mathsf{mono\_fail}.R=(\forall u,u',x:u\le u'\land R.(u,x).\bullet\impl R.(u',x).\bullet)
\\[1ex]
\mathsf{indep1}.R=(\forall x,y,z,z':R.((\bot,\bot),z).((x,y),z')\\ \ \ \  \ \ \impl
(\exists a:R.((x,\bot),z).((x,y),a))\land(\exists b:R.((\bot,y),z).((x,y),b)))\\[1ex]
\mathsf{indep2}.R=(\forall x,y,y',a,b,a',b':R.((\bot,\bot).x).((a,b),y)
\land R.((\bot,\bot).x).((a',b'),y')
\\ \ \ \ \ \ \impl(\exists z:R.((\bot,\bot).x).((a,b'),z)))
\end{array}
\]
In these definitions we assume that all variables range over proper
values only, and not $\bullet$. \textcolor{red}{{} }

The property $\monorel.R$ is true if the relation $R$ is such that
when the input $u$ contains more unknown values than input $u'$
then all choices available from $u'$ are also available from $u$.
If $\monorel.R$ is true, then $(\fba.R)^{+}=\fba.R$, and the feedback
computation becomes easier. The property $\mathsf{mono\_fail}.R$
is dual, and states that if $R$ fails for some input $u$ and some
$u'$ has less unknown values than $u$, then $R$ fails also for
$u'$. The properties $\mathsf{indep1}$, and $\text{\ensuremath{\mathsf{indep2}}}$
introduces some independences in the way the first input component
of $R$ is used in the computation.

Next theorem gives sufficient conditions under which the equality
$\fbfun.R=\seqfb.R$ holds. 
\begin{thm}
Let $R$ a relation as in Figure \ref{fig:compose-inst}(a). We
have:\end{thm}
\begin{enumerate}
\item If $\monorel.R$ is true then for all proper values (possibly unknowns,
but not fail) $x,v,t,y$ 
\[
\fbfun.R.x.((v,t),y)=\seqfb.R.x.((v,t),y)
\]

\item If $\monorel.R$, $\mathsf{mono\_fail}.R$, $\mathsf{indep1}.R$,
and $\mathsf{indep2}.R$, then for all $x$ 
\[
\fbfun.R.x.\bullet=\seqfb.R.x.\bullet
\]

\end{enumerate}

\subsection{\label{sub:rel-pt}Instantaneous feedback on predicate transformers}

In this section we show how to construct the instantaneous feedback
directly on predicate transformers.

We show how to define $\fba$, $\fbb$, $\fbbegin$, $\fbend$, $\fbfun$,
and $\fb$ on predicate transformers over values with $\bot$. 
\begin{defn}
If $S$ is a monotonic predicate transformer, then we define 
\[
\begin{array}{ll}
\Fbbegin= & [x\leadsto\bot,x]\\
\Fbend= & \assert{v,y\:|\;\maximal.v}\comp[v,y\leadsto y]\\
\Fba.S= & \assume{u,x\leadsto(u,x),x}\\
 & \quad\comp((S\ \|\ [u,x\leadsto v,y\:|\: u<v])\times\sskip)\\
 & \quad\comp[(v,y),x\leadsto(v,x)]\\
\Fbb.S= & \fusion.( S, [u,x\leadsto u,y])\\
\Fbfun.S= & \Fbbegin\comp(\Fba.S)^{*}\comp\Fbb.S\\
\Fb.S= & \Fbfun.S\comp\Fbend
\end{array}
\]
\end{defn}

Unknown ($\bot$) occurs explicitly in the definition of $\Fbbegin$,
and it occurs implicitly in the definitions of $\Fbend$ and $\Fba$
because the order $\le$ relation on values is defined using $\bot$. 
\begin{thm}
\label{thm:rel-pt}If $R$ is a relation with fail and unknown, then\end{thm}
\begin{enumerate}
\item $\wpr.(\fbbegin)=\Fbbegin$ 
\item $\wpr.(\fbend)=\Fbend$ 
\item $\wpr.(\fba.R)=\Fba.(\wpr.R)$ 
\item $\wpr.(\fbb.R)=\Fbb.(\wpr.R)$ 
\item $\wpr.(\fbfun.R)=\Fbfun.(\wpr.R)$ 
\item $\wpr.(\fb.R)=\Fb.(\wpr.R)$ 
\end{enumerate}
Parts 1 to 4 from this theorem can be proved expanding the
definitions, and using properties of fusion and product (Theorem \ref{thm:product-fusion}).
Part 5 follows from 1, 3, and 4 using $\wpr.(R^{*})=(\wpr.R)^{*}$
(Lemma \ref{lem:star}). Part 6 follows directly from 2 and 5.

Theorem \ref{thm:rel-pt},
in particular, parts 5 and 6, show that the definition of the feedback
operation on predicate transformers is equivalent to the definition
of feedback on relations with fail and unknown.
\fi

\section{Iteration Operators}
\label{sec:iteration}

In Section \ref{sec:feedback-unit-delay} we will define feedback
with unit delay for arbitrary predicate transformers. In this section
we introduce some auxiliary operators to aid that definition.

For a property transformer $S$ with input and output of the same
type we want to construct another property transformer $\iteratenextomega.S$
which works in the following way. It starts by executing $S$ on the
input sequence \textbf{$\mathbf{x}$} to obtain the output $\mathbf{y}$,
then it executes $S$ on the suffix $\mathbf{y}^{1}$ ($\mathbf{y}$ starting
from position $1$ instead of $0$) to obtain
$\mathbf{z}$, then it executes $S$ on $\mathbf{z}^{1}$, and so on.
The output of $\iterateomega.S$ is the sequence $\mathbf{y}_{0},\;\mathbf{z}_{0},\;\cdots$. 
The operator $\iteratenextomega$ is illustrated in Figure \ref{iteratenextomega}.
In the figure, we use multiple incoming arrows to represent {\em sequences} of
values, and not separate input ``wires''.

\begin{figure}[h]
\begin{centering}
\ifdefined\techrep
\includegraphics{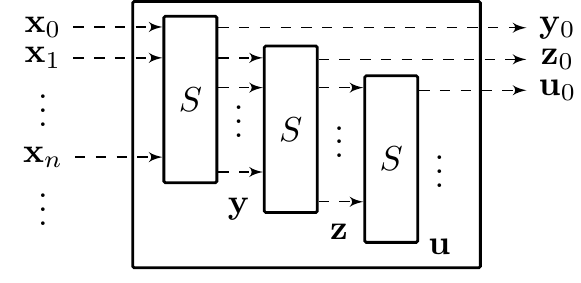} 
\else
\includegraphics[width=.6\columnwidth]{fig/iterate-next-omega} 
\fi
\par\end{centering}
\protect\protect\caption{ $\protect\iteratenextomega.S$}
\label{iteratenextomega} 
\end{figure}
The first step in defining $\iteratenextomega.S$ is to define a property
transformer $\skipnext.S$ which for some input $\mathbf{x}$, executes
$S$ on $\mathbf{x}^{1}$and leaves the first component $\mathbf{x}_{0}$
of $\mathbf{x}$ unchanged. 
\begin{defn}
If $S$ is a property transformer and $n\in\nat$, then $\skipnext.S$ and $\skiptop.n$ are property transformers given
by 
\[
\begin{array}{ll}
\skipnext.S &= [\mathbf{x} \leadsto \mathbf{x}_0, \mathbf{x}^1]
\comp (\mathsf{Skip} \times S) \comp [z,\mathbf{w}\leadsto z\cdot \mathbf{w}] \\
\skiptop.n & =[\mathbf{x}\leadsto \mathbf{y}\:|\;\eqtop.n.\mathbf{x}.\mathbf{y}]\\
\end{array}
\]
where $\eqtop.n.\mathbf{x}.\mathbf{y} = (\forall i<n:\mathbf{x}_{i}=\mathbf{y}_{i})$ and $z\cdot\mathbf{w}$ denotes the new sequence obtained by prepending
element $z$ at the head of sequence $\mathbf{w}$.
\end{defn}
Figure \ref{NextSkipNext} gives a graphical illustration of these transformers. In the figure we use double arrows to represent
sequences.

Intuitively, $\skipnext.S$ takes the sequence $\mathbf{x}$, removes its ``head''
$\mathbf{x}_0$, applies $S$, and then propends to the output the
head $\mathbf{x}_0$.
$\skiptop.n$ takes some sequence $\mathbf{x}$ and outputs a sequence
$\mathbf{y}$ which has the same first $n$ elements as $\mathbf{x}$, and is
arbitrary after that.
\begin{figure}[h]
\begin{centering}
\ifdefined\techrep
\includegraphics{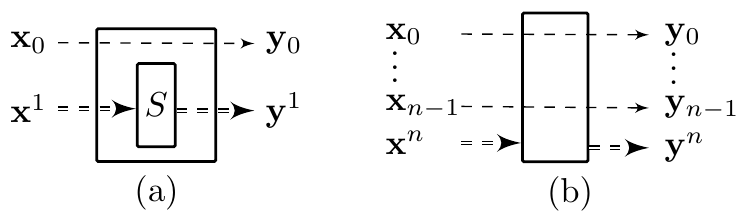} 
\else
\includegraphics[width=.84\columnwidth]{fig/skipnexttop} 
\fi
\par\end{centering}
\protect\protect\caption{(a) $\protect\skipnext.S$, (b) $\protect\skiptop.n$}
\label{NextSkipNext} 
\end{figure}

\ifdefined\techrep
$\skipnext$ and $\skiptop$ satisfy the following properties:
\begin{thm}
\label{thm:next-skipnext}If $S$, $T$ are property transformers,
$p$ is a property, and $r$ is a relation, then we have\end{thm}
\begin{enumerate}
\item $\skipnext.(S\comp T)=(\skipnext.S)\comp(\skipnext.T)$ 
\item $\skipnext.\fusion(S, T)=\fusion\big((\skipnext.S),(\skipnext.T)\big)$ 
\item $\skipnext.(\assert p\comp\assume r)=\assert{\nex p}\comp\assume{\idnext.r}$ 
\item $\skiptop.0=\havocc$ 
\item $S\sqsubseteq S'\impl\skipnext.S\sqsubseteq\skipnext.S'$ 
\end{enumerate}
where $\idnext.r=(\nex r)\land\eqtop.1$.
\fi

Then, $\iteratenextomega.S$ would intuitively be the infinite serial composition
\[
\iteratenextomega.S=S\comp\skipnext.S\comp\ldots\comp\skipnext^{n}.S\comp\ldots
\]
Formally, however, we cannot define an infinite serial composition.
In order to solve this problem, we construct the finite compositions
\[
S\comp\skipnext.S\comp\ldots\comp\skipnext^{n}.S\comp\skiptop.n
\]
for all $n$, and then we take the infinite fusion of all these property
transformers. 
\ifdefined\techrep
Formally we have: 
\begin{defn}
\label{def:iteration}If $S$ is a property transformer, $n\in\nat$,
and $F$ is a mapping on monotonic property transformers, then $\iterate.F.S.n$,
$\iterateomega.F.S$, and $\iteratenextomega.S$ are defined by
\begin{equation}
\begin{array}{lll}
\iterate.F.S.n & =& S\comp F.S\comp\ldots\comp F^{n-1}.S\\
\iterateomega.F.S & =& \fusion.(\lambda n:\iterate.F.S.n\comp\skiptop.n) \\
\iteratenextomega.S & = & \iterateomega.\skipnext.S
\end{array}\label{eq:iteration}
\end{equation}
\end{defn}
The property transformer $\iterate.\skipnext.S.n$ applied to a sequence
$\mathbf{x}$ calculates a sequence $\mathbf{y}$ where the first
$n$ components are correct values for a possible output of $\iteratenextomega.S$
as can be seen in Figure \ref{iteratenextomega}. Always the first
$n$ components of $\mathbf{y}$ are possible values computed by $\iteratenextomega.S$,
however the components $\mathbf{y}^{n}$ of $\mathbf{y}$ are not
correct components for the final result of $\iteratenextomega.S$.
For this reason we set these components to arbitrary values using
$\skiptop.n$. Using $\iterate.\skipnext.S.n\comp\skiptop.n$ for
arbitrarily large $n$s we get arbitrary large prefixes of outputs
of $\iteratenextomega.S$. Taking the fusion of all these sequences together
results in all possible outputs of $\iteratenextomega.S$. 

\begin{example}[Iterate of unit delay]
Let us apply $\iterate.\skipnext$ to a unit delay initialized by the value
$a$. If $\delay.a=[x\leadsto y\;|\; y_{0}=a\land(\forall i.y_{i+1}=x_{i})]$,
then we have 
\begin{equation}
\begin{array}{l}
\iterate.\skipnext.(\delay.0).n\comp\skiptop.n
\;\;=\;\;[\mathbf{x}\leadsto\mathbf{y}\;|\;\forall i<n:\mathbf{y}_{i}=0]
\end{array}\label{eq:unitdelay}
\end{equation}
that is, from arbitrary input $\mathbf{x}$ we get all possible $\mathbf{y}$s
that have a prefix of size $n$ of $0$s. 
\ifdefined\techrep
The proof of (\ref{eq:unitdelay}) is given by:
\begin{lyxlist}{00}
\item [{~}] $\iterateomega.\skipnext.(\delay.0)$ 
\item [{$=$}] \{Definition of $\iterateomega$ and (\ref{eq:unitdelay})\} 
\item [{~}] $\fusion.(\lambda n:[\mathbf{x}\leadsto\mathbf{y}\;|\;\forall i<n:\mathbf{y}_{i}=0])$ 
\item [{$=$}] \{Theorem \ref{thm:product-fusion}\} 
\item [{~}] $[\mathbf{x}\leadsto\mathbf{y}\;|\;(\forall n:\forall i<n:\mathbf{y}_{i}=0]$ 
\item [{$=$}] \{Logic calculation\} 
\item [{~}] $[\mathbf{x}\leadsto\mathbf{y}\;|\;\forall i:\mathbf{y}_{i}=0]$ 
\end{lyxlist}
\end{example}
\fi
\ifdefined\techrep
We now state some properties for the iteration operators introduced
in Definition \ref{def:iteration}. The iterations preserve refinement. 
\begin{thm}
If $F$ is a monotonic mapping of property transformers, then\end{thm}
\begin{enumerate}
\item $S\sqsubseteq S'\impl\iterate.F.S.n\sqsubseteq\iterate.F.S'.n$ 
\item $S\sqsubseteq S'\impl\iterateomega.F.S\sqsubseteq\iterateomega.F.S'$ 
\item $S\sqsubseteq S'\impl\iteratenextomega.S\sqsubseteq\iteratenextomega.S'$ 
\end{enumerate}
\fi

\ifdefined\techrep
The next two theorems show how to calculate the iteration operators for
property transformers of the form $\assert p\comp\assume r$. Before
stating these theorems we introduce an iteration operator on relations,
similar to the one on monotonic transformers.
\else
\fi
For a relation $r:X\to X\to\bool$
and a mapping $R:(X\to X\to\bool)\to(X\to X\to\bool)$ on relation
we define 
\[
\iterel.R.r.n=r\circ R.r\circ\cdots\circ R^{n-1}.r.
\]

\ifdefined\techrep
\begin{thm}
If $p$ is a predicate, $r$ a relation, $F$ a mapping on monotonic
transformers, $P$ a mapping on predicates, and $R$ a mapping on
relations, then\end{thm}
\begin{enumerate}
\item If $(\forall p,r:F.(\{p\}\comp[r])=\{P.p\}\comp[R.r])$ then\\
 $\iterate.F.(\{p\}\comp[r]).n \;\;=\;\;
 \{\mathbf{x}\;|\;\forall i<n:\forall y:\iterel.R.r.i.\mathbf{x}.\mathbf{y}\impl P^{i}.p.\mathbf{y}\}\comp [\iterel.R.r.n]$ 
\item $\iterate.\skipnext.(\{p\}\comp[r]).n \;=\;
 \{\mathbf{x}\,|\,\forall i<n:\forall y:\iterel.\idnext.r.i.\mathbf{x}.\mathbf{y}\impl\nex^{i}.p.y\}\comp [\iterel.\idnext.r.n]$ 
\end{enumerate}
Using this theorem, Theorem \ref{thm:product-fusion}, and Theorem
\ref{thm:next-skipnext} we can calculate 
 $\iteratenextomega.(\{p\}\comp[r])$. 
\else
 The next theorem shows how to calculate $\iteratenextomega.(\{p\}\comp[r])$. 
\fi
\begin{thm}
For a property $p.\mathbf{x}$ and a relation $r.\mathbf{x}.\mathbf{y}$
we have 
\[
\begin{array}{l}
\iteratenextomega.(\{p\}\comp[r])
\;\;=\;\;
\{\mathbf{x}\;|\;(\forall n,\mathbf{y}:\iterel.\idnext.r.n.\mathbf{x}.\mathbf{y}\impl\nex^{n}p.\mathbf{y})\}\\
\hspace{5cm}\comp[\,\bigwedge_{n}(\iterel.\idnext.r.n\circ\eqtop.n)\,]
\end{array}
\]
\ifdefined\techrep
\else
where $\idnext.r=(\nex r)\land\eqtop.1$.
\fi
\end{thm}
\else
The details are omitted due to lack of space, and can be found in the
extended version of this paper~\cite{preoteasa:tripakis:2015}.
In~\cite{preoteasa:tripakis:2015} we also show how to calculate
$\iteratenextomega.S$ for the special case when the property transformer $S$
is of the form $\{p\}\comp[r]$. This result is used to prove
Theorem~\ref{thm:delay-feedback} in the section that follows.
\fi

\section{\label{sec:feedback-unit-delay}Feedback with Unit Delay}
In Section~\ref{sec:inst-feedback} we defined an instantaneous feedback operator, for stateless systems. In this section we turn our attention to feedback for systems with state.
We model a stateful system as a predicate transformer $S$ with inputs 
$u$ and $x$ and outputs $v$ and $y$, such that $u$ models the {\em current}
state, and $v$ the {\em next} state (thus, $u$ and $v$ must have the same type).
We will connect $v$ to $u$ in feedback via a unit delay. By doing so, the next
state becomes the current state in the next step.

The result of the feedback-with-unit-delay operator applied to a predicate
transformer as above is a property transformer that
takes as input an infinite sequence $\mathbf{x}$ and produces an
infinite output sequence $\mathbf{y}$,
for a given initial state $\mathbf{u}_{0}$. The system starts by executing
$S$ on the first component $\mathbf{x}_{0}$ of $\mathbf{x}$, and
on $\mathbf{u}_{0}$. This results in the values
$\mathbf{u}_{1}$ and $\mathbf{y}_{0}$. In the next step, the second
value $\mathbf{x}_{1}$of $\mathbf{x}$ and $\mathbf{u}_{1}$ are
used as input for $S$ resulting in $\mathbf{u}_{2}$ and $\mathbf{y}_{1}$,
and so on. This computation is depicted in Figure \ref{unitdelayfeedback}.

We achieve this computation by lifting $S$ to a special property
transformer and then applying the $\iteratenextomega$ operator. 
\begin{defn}
For a predicate transformer $S$ as described earlier, we define the
property transformer $\adddelay.S$ with input a tuple of infinite
sequences $\mathbf{u},\mathbf{x},\mathbf{y}$ and output of the same
type: 
$$
\begin{array}{l}
\hspace{-.3cm}
\adddelay.S= [\mathbf{u},\mathbf{x},\mathbf{y}\leadsto
(\mathbf{u}_{0},\mathbf{x}_{0}), (\mathbf{u}_0,\mathbf{x})]\comp (S\times \mathsf{Skip}) \comp \\
\hspace{-.2cm}
\mbox{}[(a,b),(u,\mathbf{x})\leadsto \mathbf{u'},\mathbf{x'},\mathbf{y'}
 | \mathbf{u'}_0 = u \land \mathbf{u'}_1 = a 
\land\mathbf{y'}_0 = b \land \mathbf{x'} = \mathbf{x}]
\end{array}
$$
\end{defn}
The property transformer $\adddelay.S$ takes as input the sequences
$\mathbf{u},\mathbf{x},\mathbf{y}$, executes $S$ on input $\mathbf{u}_{0}$
and $\mathbf{x}_{0}$ producing the values $a$, and $b$, and then
sets the output $\mathbf{u}',\mathbf{x}',\mathbf{y}'$ where $\mathbf{u}'_{0}=\mathbf{u}_{0}$,
$\mathbf{u}'_{1}=a$, $\mathbf{y}'_{0}=b$, and $\mathbf{x}'=\mathbf{x}$.
The components $\mathbf{u}'_{2},\mathbf{u}'_{3},\ldots$ and $\mathbf{y}'_{1},\mathbf{y}'_{2},\ldots$
are assigned arbitrary values. The transformer $\adddelay.S$ returns
the value of the second output component $y$ at time zero, but adds
a unit delay to the first output component $v$. $\adddelay.S$ also
adapts the input and the output of $S$ to be of the same type such
that we can apply $\iteratenextomega$ to $\adddelay.S$. Figure
\ref{adddelay} gives a graphical representation of $\adddelay.S$.
In this figure the input $\mathbf{x}$ is split in two components
$\mathbf{x}_{0}$ and $\mathbf{x}^{1}=(\mathbf{x}_{1},\mathbf{x}_{2},\ldots)$,
input $\mathbf{u}$ is split similarly in $\mathbf{u}_{0}$ and $\mathbf{u}^{1}$,
output $\mathbf{u}'$ is split in $\mathbf{u}'_{0}$, $\mathbf{u}'_{1}$,
and $\mathbf{u}'^{2}$, output $\mathbf{x}'$ is equal to input $\mathbf{x}$,
and output $\mathbf{y}'$ is split in $\mathbf{y}_{0}$ and $\mathbf{y}^{1}$.
From this figure we can also see that the input components $\mathbf{u}^{1}$
and $\mathbf{y}$ are not used (they are not connected inside $\adddelay.S$),
and the output components $\mathbf{u}'^{2}$ and $\mathbf{y}'^{1}$
are assigned arbitrary values (they are not connected).

\begin{figure}[h]
\centering{}
\ifdefined\techrep
	\subfloat[]{\label{unitdelayfeedback}
		\includegraphics{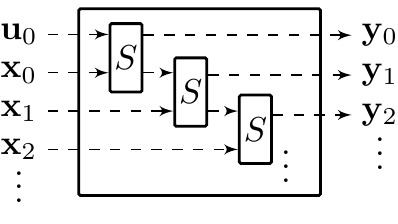}}	\ \ \ \ \ 
	\subfloat[]{\label{adddelay}
		\includegraphics{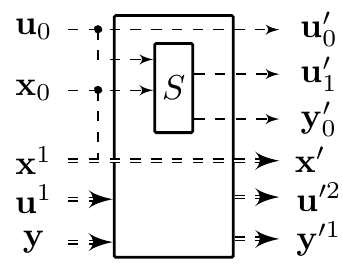}}	
\else
	\subfloat[]{\label{unitdelayfeedback}
		\includegraphics[width=.4\columnwidth]{fig/feedback}}	\ \ \ \ \ 
	\subfloat[]{\label{adddelay}
		\includegraphics[width=.3\columnwidth]{fig/unit}}	
\fi
\protect\protect\caption{(a) Unit delay feedback of $S$, (b) $\mathsf{AddDelay}.S$}
\end{figure}

Finally we can introduce the transformer $\delayfeedback.init.S$: 
\begin{defn}
If $init$ is a predicate on the initial value of the feedback variable
$u$ and $S$ is a predicate transformer as described above, then
unit delay feedback is given by 
\[
\begin{array}{l}
\delayfeedback.init.S=[\mathbf{x}\leadsto\mathbf{u},\mathbf{x},\mathbf{y}\;|\; init.\mathbf{u}_{0}]\comp
\ifdefined\techrep\else\\ \qquad\quad\fi
\iteratenextomega(\adddelay.S)\comp[\mathbf{u},\mathbf{x},\mathbf{y}\leadsto\mathbf{y}]
\end{array}
\]
\end{defn}

The intuition behind the above definition is that if we plug $\adddelay.S$
(Figure \ref{adddelay}) into Figure \ref{iteratenextomega}, using
the tuple $(\mathbf{u},\mathbf{x},\mathbf{y})$ instead of $\mathbf{x}$,
then we obtain the desired system of Figure \ref{unitdelayfeedback}.

A major result of this section is that the operators
$\adddelay$ and $\delayfeedback$ preserve refinement:
\begin{thm}
\label{thm:refin-delay}If $S$ is a predicate transformer as above
then\end{thm}
\begin{enumerate}
\item $S\sqsubseteq S'\impl\adddelay.S\sqsubseteq\adddelay.S'$ 
\item $S\sqsubseteq S'\impl\delayfeedback.S\sqsubseteq\delayfeedback.S'$ 
\end{enumerate}

The results presented so far in this section apply to any predicate 
transformer $S$
with the structure described above (inputs $u,x$ and outputs $v,y$).
In practice, we often describe transformers as symbolic transition systems,
that is, using a predicate $init$ specifying the set of initial states,
a predicate $p$ specifying the set of legal inputs, and a relation $r$
specifying the transition and output relations.
If $S$ is obtained from a symbolic transition system, that is, if
$S$ has the form 
$\{u,x\;|\; p.u.x\}\comp$ $[u,x\leadsto u',y\;|\; r.u.u'.x.y]$, then
we can prove the following:
\begin{thm} We have
\label{thm:delay-feedback}
\[
\begin{array}{l}
\delayfeedback.init.(\{u,x\,|\, p.u.x\};[u,x\leadsto u',y\,|\, r.u.u'.x.y])
\ifdefined\techrep\else\\\fi
\;=\;
\assert{\mathbf{x}\;|\;\forall\mathbf{u}:init.\mathbf{u}_{0}\impl(\inp.r\leads p).\mathbf{u}.\mathbf{u}^{1}.\mathbf{x}}\\
\ifdefined\techrep\hspace{9.5cm}\else\qquad\quad\fi
\comp\assume{\mathbf{x}\leadsto\mathbf{y}\;|\;\exists\mathbf{u}:init.\mathbf{u}_{0}\land\Box\, r.\mathbf{u}.\mathbf{u}^{1}.\mathbf{x}.\mathbf{y}}
\end{array}
\]
\end{thm}

This theorem shows that the unit delay feedback of such a predicate
transformer is identical to the property transformer semantics of the 
corresponding symbolic transition system~\cite{preoteasa:tripakis:2014},
which provides a sanity check of our construction.
The
condition $(\forall\mathbf{u}:init.\mathbf{u}_{0}\impl(\inp.r\leads p).\mathbf{u}.\mathbf{u}^{1}.\mathbf{x})$
is false on input $\mathbf{x}$, if there exists $\mathbf{u}_{0},\ldots,\mathbf{u}_{n}$
and $\mathbf{y}_{0},\ldots,\mathbf{y}_{n-1}$ such that $(\forall i<n:r.\mathbf{u}_{i}.\mathbf{u}_{i+1},\mathbf{x}_{i},\mathbf{y}_{i})$
is true and $p.\mathbf{u}_{n}.\mathbf{x}_{n}$ is false. That is,
if with $\mathbf{x}$ it is possible to reach a false assertion, then
the system will fail from $\mathbf{x}$, even if non-failing choices
are possible. 

The intuitive behavior of a system as in Theorem~\ref{thm:delay-feedback}
was explained in the beginning of this section.
Here we elaborate on this intuition further, explaining 
in particular how it works for non-deterministic and non-input-receptive systems.
 The system starts with
some initial state $\mathbf{u}_{0}$ and input $\mathbf{x}_{0}$
(the first element of $\mathbf{x}$)
and it checks the input condition $p.\mathbf{u}_{0}.\mathbf{x}_{0}$.
If the condition is false, then $\mathbf{x}_{0}$ is an illegal input
at state $\mathbf{u}_{0}$, and execution fails. Otherwise, the system
computes non-deterministically the output $\mathbf{y}_{0}$ and the
next state $\mathbf{u}_{1}$,
such that $r.\mathbf{u}_{0}.\mathbf{u}_{1}.\mathbf{x}_{0}.\mathbf{y}_{0}$
is true. This ends the first execution step. The system then proceeds
with the next step, where it uses $\mathbf{u}_{1}$
and $\mathbf{x}_{1}$ to test again the input condition $p$, and so on.
At any point during execution, if $p.\mathbf{u}_{i}.\mathbf{x}_{i}$ becomes
false then the system fails. Otherwise the system outputs non-deterministically
the infinite sequence $\mathbf{y}_{0},\mathbf{y}_{1},\cdots$. 
We note that if $r$ is deterministic, i.e., if there exists function $f$ such that $r.u.u'.x.y \Leftrightarrow (u',y)=f.(u,x)$, then this process yields the standard execution runs.

\begin{example}\label{ex:sum:delay}[Stateful systems in feedback]
Consider the following relations modeling systems with
state ($u$ and $u'$ represent current and next state variables, respectively):
\begin{eqnarray}
R_1 & = & [u,x\leadsto u',y\;|\; u'=u\land y=u] \\
R_2 & = & [u,x\leadsto u',y\;|\; u'=u+1\land y=u]
\end{eqnarray}
In $R_1$, the next state is equal to the current state.
In $R_2$, the next state increments the current state by $1$.
We will also use the relation $\mathit{StepSum}$ defined in~(\ref{eqstepsum}).
In all three cases the output $y$ is equal to the state.
Let $\mathsf{Zero}$ be the predicate given by $\mathsf{Zero}.u=(u=0)$. 
We compose the above systems by connecting $u'$ to $u$ in feedback
with unit delay.
In all three cases we use the predicate $\mathsf{Zero}$ as $init$,
and in all three cases we use input-receptive
components, i.e., where $p=\top$.
We get:
\begin{enumerate}
\item $\delayfeedback.\mathsf{Zero}.R_1=[\mathbf{x}\leadsto\mathbf{y}\,|\,\forall i:\mathbf{y}_{i}=0]$.
Since the initial state is $0$, the output
is always $0$. 
\item $\delayfeedback.\mathsf{Zero}.R_2=[\mathbf{x}\leadsto\mathbf{y}\,|\,\forall i:\mathbf{y}_{i}=i]$.
The output at the $i$-th step equals $i$. 
\item $\delayfeedback.\mathsf{Zero}.\mathit{StepSum}=[\mathbf{x}\leadsto sum.\mathbf{x}]$.
The output is the summation of the sequence of inputs seen so far. 
\end{enumerate}
\end{example}

The three examples above are all input-receptive and deterministic.
Our framework can also handle non-input-receptive
and non-deterministic systems, as illustrated next:
\begin{example}[A deterministic but non-input-receptive system]
Let $R_3 = \assert{u,x\;|\; u\le a}\comp
[u,x\leadsto u',y\;|\; u'=u+x\land y=u]$.
$R_3$ is similar to $\mathit{StepSum}$, but imposes the condition that at each step
the state $u$ is bounded by some constant $a$.
Applying feedback we get:
\[
\delayfeedback.\mathsf{Zero}.R_3
=\{\mathbf{x}|\forall i:sum.\mathbf{x}.i\le a\}\comp[\mathbf{x}\leadsto sum.\mathbf{x}]
\]
Note that the local (i.e., at each step) condition $u\le a$ becomes (after
feedback) the global condition that the sum of all elements $\mathbf{x}_{i}$
of every prefix of the input $\mathbf{x}$ must be less or equal to
$a$. 
\end{example}

\begin{example}[A non-deterministic and non-input-receptive system]
We can also weaken the relation $u'=u+x\land y=u$ and make the previous
system non-deterministic.
For example at each step we can choose to compute $u'=u+x$ or $u'=x$.
Let $R_4=\assert{u,x\;|\; u\le a}\comp [u,x\leadsto u',y|(u'= u+ x\lor u'= x)\land y=u]$.
Then we get: 
\[
\begin{array}{l}
\delayfeedback.\mathsf{Zero}.R_4
=\{\mathbf{x}\;|\;\forall i:sum.\mathbf{x}.i\le a\}\comp
\ifdefined\techrep\else\\ \qquad\fi
[\mathbf{x}\leadsto\mathbf{y}\;|\;\mathbf{y}_{0}=0\land(\forall i:\mathbf{y}_{i+1}=\mathbf{y}_{i}+\mathbf{x}_{i}\lor\mathbf{y}_{i+1}=\mathbf{x}_{i})]
\end{array}
\]
Observe that the global condition on $\mathbf{x}$ remains the same.
This is because in the worst case the non-deterministic choice could
always pick the alternative $u'=u+x$, which leads to higher numbers
stored in $u$ compared to the alternative $u'=x$. 
\end{example}

\ifdefined\techrep
\section{Symbolic Computation}
\label{sec:algorithm}

For a relation $R:((A_{1\bot}\times\ldots\times A_{k\bot})\times X)^{\bullet}\to((A_{1\bot}\times\ldots\times A_{k\bot})\times Y)^{\bullet}$,
in the $\exists n$ quantifier from the definition of $\fbfun.R$,
$n$ is bound by $k$. This is so because a strictly increasing sequence
$u_{0}<u_{1}<\ldots$ on $A_{1\bot}\times\ldots\times A_{k\bot}$\textcolor{black}{{}
can have at most $k+1$ elements. Because of this property we have:}
\[
(\fba.R)^{*}=\bigvee_{i\le k}(\fba.R)^{i}
\]
so we can calculate the feedback symbolically using 
\[
\fbfun.R=\fbbegin\circ(\bigvee_{i\le k}(\fba.R)^{i})\circ\fbb.R
\]

\sloppypar{If relation $R$ is defined using a first order formula, then $\fbfun.R$ is also a first order formula.}

Checking refinement of two relations with $R\sqsubseteq R'$ defined
by first order formulas is also reduced to checking validity of a
first order formula $(\forall x:R.x.\bullet\lor(\forall y:R'.x.y\impl R.x.y)$.

Refinement of predicate transformers of the form $\{p\}\comp[r]$
are also reduced to checking validity of a first order formula (Theorem
\ref{thm:spec-comp-refin}). The predicate transformers of the form
$\{p\}\comp[r]$ are closed to serial and demonic compositions, fusion,
and product, so if $p$ and $r$ are defined using first order formulas,
then so are the results of these operations (Theorem \ref{thm:spec-comp-refin}
and \ref{thm:product-fusion}).

The delay feedback of a predicate transformer of the form $\{p\}\comp[r]$
is equivalent by Theorem \ref{thm:delay-feedback} to a property transformer
of the form $\{P\}\comp[R]$ where $P$ and $R$ are quantified LTL
formulas. Checking the refinement $\delayfeedback.S\sqsubseteq\delayfeedback.S'$,
is reduced by Theorem \ref{thm:refin-delay} to checking the refinement
$S\sqsubseteq S'$ of predicate transformers.

However, $\iterateomega.F.S$ and $\iteratenextomega.S$ cannot be
reduced always to first order logic or quantified LTL formulas.
\fi

\section{Implementation in Isabelle}

We have stated and proven all results presented in this paper,
\ifdefined\techrep
\else
as well as in the extended version~\cite{preoteasa:tripakis:2015},
\fi
in the Isabelle/HOL theorem prover. 
Our theory is built on top of theories for refinement
calculus and a semantic formalization of LTL. Proofs omitted 
from this paper can be found in the Isabelle code,
available from 
\url{http://rcrs.cs.aalto.fi}.
At the time of this writing our Isabelle theories comprise a total
of about 5k lines of code.

\ifdefined\techrep
Our formalization benefited from the rich collection of Isabelle
formalization of mathematical structures used in this paper: partial
orders, Boolean algebras, complete lattices, and others.
\fi

\section{Conclusions}

Feedback composition is challenging to define for non-deterministic
and non-input-receptive systems, so that refinement is preserved.
In this paper we make progress toward this goal. 
First, we propose an instantaneous feedback composition operator for
partial relations with fail and unknown (Section~\ref{sec:inst-feedback}). 
This operator generalizes constructive semantics which are limited to total
functions~\cite{Malik94,Berry99,EdwardsLee03}.
Second, we propose a feedback-with-unit-delay operator for systems with state
(Section~\ref{sec:feedback-unit-delay}).
Both operators preserve refinement
(Theorems~\ref{thm:feedback:refinement} and~\ref{thm:refin-delay}).

Our framework can be applied to arbitrary Simulink diagrams by
first translating them into predicate transformers that handle current
and next state variables as inputs and outputs, respectively,
and then applying feedback with unit delay on these variables.
Contrary to Simulink,
our construction allows instantaneous feedback as well as non-deterministic
and non-input-receptive components.

Open problems remain. Seeking a generalization of the feedback operators
directly to property transformers, at the semantic level, is a worthwhile
albeit difficult goal. With this generalization, we will be able to
treat Simulink basic blocks as property transformers, and apply serial
and feedback compositions directly.

\ifdefined\techrep
\bibliographystyle{alpha}
\else
\bibliographystyle{abbrvnat}
\fi
\bibliography{bibl}

\newcommand{\etalchar}[1]{$^{#1}$}
\begin{thebibliography}{JDK{\etalchar{+}}14b}

\bibitem[AH99]{AlurHenzingerFMSD99}
R.~Alur and T.~Henzinger.
\newblock Reactive modules.
\newblock {\em Formal Methods in System Design}, 15:7--48, 1999.

\bibitem[BB95]{back:butler:1995}
R.-J. Back and M.~Butler.
\newblock Exploring summation and product operators in the refinement calculus.
\newblock In {\em Mathematics of Program Construction}, volume 947 of {\em
  LNCS}. Springer, 1995.

\bibitem[Ber99]{Berry99}
G.~Berry.
\newblock {The Constructive Semantics of Pure Esterel}, 1999.

\bibitem[BS01]{BroyStolen01}
M.~Broy and K.~St{\o}len.
\newblock {\em Specification and development of interactive systems: focus on
  streams, interfaces, and refinement}.
\newblock Springer, 2001.

\bibitem[BvW98]{back-wright-98}
R.-J. Back and J.~von Wright.
\newblock {\em Refinement Calculus. A systematic Introduction}.
\newblock Springer, 1998.

\bibitem[dA04]{AlfaroGamesOpenSystems04}
L.~de~Alfaro.
\newblock Game models for open systems.
\newblock In Nachum Dershowitz, editor, {\em Verification: Theory and
  Practice}, volume 2772 of {\em Lecture Notes in Computer Science}, pages
  192--213. Springer, 2004.

\bibitem[dAH01]{AlfaroHenzingerFSE01}
L.~de~Alfaro and T.~Henzinger.
\newblock Interface automata.
\newblock In {\em Foundations of Software Engineering (FSE)}. ACM Press, 2001.

\bibitem[DHJP08]{DoyenHJP08}
L.~Doyen, T.~Henzinger, B.~Jobstmann, and T.~Petrov.
\newblock Interface theories with component reuse.
\newblock In {\em EMSOFT}, pages 79--88, 2008.

\bibitem[Dij75]{dijkstra-75}
E.W. Dijkstra.
\newblock Guarded commands, nondeterminacy and formal derivation of programs.
\newblock {\em Comm. ACM}, 18(8):453--457, 1975.

\bibitem[DP02]{priestley}
B.A. Davey and H.A. Priestley.
\newblock {\em Introduction to lattices and order}.
\newblock Cambridge University Press, New York, second edition, 2002.

\bibitem[EL03]{EdwardsLee03}
S.~Edwards and E.A. Lee.
\newblock The semantics and execution of a synchronous block-diagram language.
\newblock {\em Sci. Comp. Progr.}, 48:21--42(22), July 2003.

\bibitem[FP91]{Freeman:Pfenning:1991}
T.~Freeman and F.~Pfenning.
\newblock {Refinement Types for ML}.
\newblock {\em SIGPLAN Not.}, 26(6):268--277, May 1991.

\bibitem[JDK{\etalchar{+}}14a]{jin2014ARCH}
X.~Jin, J.~Deshmukh, J.~Kapinski, K.~Ueda, and K.~Butts.
\newblock Benchmarks for model transformations and conformance checking.
\newblock In {\em 1st Intl. Workshop on Applied Verification for Continuous and
  Hybrid Systems (ARCH)}, 2014.
\newblock Benchmark Simulink models available
  from~\myhref{http://cps-vo.org/group/ARCH/benchmarks}.

\bibitem[JDK{\etalchar{+}}14b]{JinDKUB14}
X.~Jin, J.~V. Deshmukh, J.~Kapinski, K.~Ueda, and K.~Butts.
\newblock Powertrain control verification benchmark.
\newblock In {\em 17th Intl. Conf. on Hybrid Systems: Computation and Control},
  HSCC '14, pages 253--262. ACM, 2014.

\bibitem[Jon94]{Jonsson94fully}
B.~Jonsson.
\newblock A fully abstract trace model for dataflow and asynchronous networks.
\newblock {\em Distrib. Comput.}, 7(4):197--212, 1994.

\bibitem[Kah74]{Kahn74}
G.~Kahn.
\newblock The semantics of a simple language for parallel programming.
\newblock In {\em Information Processing 74, Proceedings of IFIP Congress 74}.
  North-Holland, 1974.

\bibitem[LT89]{LynchTuttle89}
N.A. Lynch and M.R. Tuttle.
\newblock An introduction to input/output automata.
\newblock {\em CWI Quarterly}, 2:219--246, 1989.

\bibitem[Mal94]{Malik94}
S.~Malik.
\newblock Analysis of cyclic combinational circuits.
\newblock {\em IEEE Trans. Computer-Aided Design}, 13(7):950--956, 1994.

\bibitem[Mil90]{millen}
J.~K. Millen.
\newblock Hookup security for synchronous machines.
\newblock In {\em Computer Security Foundations Workshop III}, pages 84--90,
  1990.

\bibitem[MTH90]{Milner:1990}
R.~Milner, M.~Tofte, and R.~Harper.
\newblock {\em The Definition of Standard ML}.
\newblock MIT Press, Cambridge, MA, USA, 1990.

\bibitem[NPW02]{nipkow-paulson-wenzel-02}
T.~Nipkow, L.~C. Paulson, and M.~Wenzel.
\newblock {\em {Isabelle/HOL} --- A Proof Assistant for Higher-Order Logic}.
\newblock LNCS 2283. Springer, 2002.

\bibitem[Plo76]{plotkin:1976}
G.~D. Plotkin.
\newblock A powerdomain construction.
\newblock {\em SIAM Journal on Computing}, 5(3):452--487, 1976.

\bibitem[Pnu77]{pnueli:1977}
A.~Pnueli.
\newblock The temporal logic of programs.
\newblock In {\em FOCS}, 1977.

\bibitem[Pre14]{preoteasa:2014}
V.~Preoteasa.
\newblock Formalization of refinement calculus for reactive systems.
\newblock {\em Archive of Formal Proofs}, October 2014.
\newblock \url{http://afp.sf.net/entries/RefinementReactive.shtml}.

\bibitem[PT14]{preoteasa:tripakis:2014}
V.~Preoteasa and S.~Tripakis.
\newblock Refinement calculus of reactive systems.
\newblock In {\em Embedded Software (EMSOFT)}. ACM, 2014.

\bibitem[RKJ08]{Rondon:Kawaguci:Ranjit:2008}
P.~M. Rondon, M.~Kawaguci, and R.~Jhala.
\newblock Liquid types.
\newblock {\em SIGPLAN Not.}, 43(6):159--169, 2008.

\bibitem[Tar55]{tarski-55}
A.~Tarski.
\newblock A lattice-theoretical fixpoint theorem and its applications.
\newblock {\em Pacific J. Math.}, 5:285--309, 1955.

\bibitem[TLHL11]{TripakisTOPLAS2011}
S.~Tripakis, B.~Lickly, T.~A. Henzinger, and E.~A. Lee.
\newblock {\ifdefined\akaprojectsep{\bf A Theory of Synchronous Relational
  Interfaces}\else A Theory of Synchronous Relational Interfaces\fi}.
\newblock {\em ACM Transactions on Programming Languages and Systems (TOPLAS)},
  33(4), July 2011.

\bibitem[TS14]{tripakisShaverFPS2014}
S.~Tripakis and C.~Shaver.
\newblock {Feedback in Synchronous Relational Interfaces}.
\newblock In {\em From Programs to Systems}, volume 8415 of {\em LNCS}, pages
  249--266. Springer, 2014.

\bibitem[XP99]{Xi:Pfenning:1999}
H.~Xi and F.~Pfenning.
\newblock Dependent types in practical programming.
\newblock In {\em POPL}, pages 214--227. ACM, 1999.

\end{thebibliography}

\end{document}